\documentclass[12pt]{article}
\usepackage{etex} 
\usepackage[greek,english]{babel}

\usepackage{enumerate}
\usepackage[all]{xy} 
\usepackage{url}
\usepackage{hyperref}

\hypersetup{
	colorlinks=true,
    linktoc=all,
    linkcolor=blue,  
    }

\renewcommand{\baselinestretch}{1.1}

\let\s=\sigma    \let\c=\chi

\newcommand{\be}{\begin{equation}}
\newcommand{\ee}{\end{equation}}
\def\ba{\begin{align}} 
\def\ea{\end{align}}
\newcommand{\bea}{\begin{eqnarray}}
\newcommand{\eea}{\end{eqnarray}}
\newcommand{\bx}{\begin{example}}
\newcommand{\ex}{\end{example}}
\newcommand{\bex}{\begin{exercise}}
\newcommand{\eex}{\end{exercise}}
\newcommand{\ban}{\begin{answer}}
\newcommand{\ean}{\end{answer}}
\newcommand{\et}{\end{theorem}}
\newcommand{\bc}{\begin{corollary}}
\newcommand{\ec}{\end{corollary}}
\newcommand{\blem}{\begin{lemma}}
\newcommand{\elem}{\end{lemma}}
\newcommand{\bp}{\begin{problem}}
\newcommand{\ep}{\end{problem}}
\newcommand{\bn}{\begin{proposition}}
\newcommand{\en}{\end{proposition}}
\newcommand{\bd}{\begin{definition}}
\newcommand{\ed}{\end{definition}}
\newcommand{\bcon}{\begin{construction}}
\newcommand{\econ}{\end{construction}}
\newcommand{\bq}{\begin{question}}
\newcommand{\eq}{\end{question}}
\newcommand{\bprf}{\begin{proof}}
\newcommand{\eprf}{\end{proof}}
\newcommand{\br}{\begin{remark}}
\newcommand{\er}{\end{remark}}
\newcommand{\bs}{\begin{solution}}
\newcommand{\es}{\end{solution}}
\newcommand{\beqs}{\begin{eqnarray}}
\newcommand{\eeqs}{\end{eqnarray}}

 \let\ov=\overline

\newcommand{\Tr}{{\rm Tr} }

\newcommand{\half}{\frac{1}{2}}

\usepackage{mathrsfs}
\usepackage[T1]{fontenc}
\usepackage{mathpazo}
\usepackage{setspace}
\usepackage{amsfonts}
\usepackage{amssymb}
\usepackage{amsmath}
\usepackage{epsfig}
\usepackage{latexsym}
\usepackage{color}
\usepackage{graphicx}
\usepackage[latin1]{inputenc}
\usepackage{slashed}
\usepackage{multirow}
\usepackage{comment}
\usepackage{soul}
\usepackage{hyperref}
\usepackage{cite}

\usepackage[titletoc]{appendix}

\def\hybrid{\topmargin -20pt    \oddsidemargin 0pt
        \headheight 0pt \headsep 0pt
        \textwidth 6.25in       
        \textheight 9.25in       
        \marginparwidth .875in
        \parskip 5pt plus 1pt   \jot = 1.5ex}

\hybrid

\def\baselinestretch{1.2}

\catcode`\@=11

\def\marginnote#1{}
\newcount\hour
\newcount\minute
\newtoks\amorpm
\hour=\time\divide\hour by60
\minute=\time{\multiply\hour by60 \global\advance\minute by-\hour}
\edef\standardtime{{\ifnum\hour<12 \global\amorpm={am}%
        \else\global\amorpm={pm}\advance\hour by-12 \fi
        \ifnum\hour=0 \hour=12 \fi
        \number\hour:\ifnum\minute<10 0\fi\number\minute\the\amorpm}}
\edef\militarytime{\number\hour:\ifnum\minute<10 0\fi\number\minute}

\def\draftlabel#1{{\@bsphack\if@filesw {\let\thepage\relax
   \xdef\@gtempa{\write\@auxout{\string
      \newlabel{#1}{{\@currentlabel}{\thepage}}}}}\@gtempa
   \if@nobreak \ifvmode\nobreak\fi\fi\fi\@esphack}
        \gdef\@eqnlabel{#1}}
\def\@eqnlabel{}
\def\@vacuum{}
\def\draftmarginnote#1{\marginpar{\raggedright\scriptsize\tt#1}}

\def\draft{\oddsidemargin -.5truein
        \def\@oddfoot{\sl preliminary draft \hfil
        \rm\thepage\hfil\sl\today\quad\militarytime}
        \let\@evenfoot\@oddfoot \overfullrule 3pt
        \let\label=\draftlabel
        \let\marginnote=\draftmarginnote
   \def\@eqnnum{(\theequation)\rlap{\kern\marginparsep\tt\@eqnlabel}%
\global\let\@eqnlabel\@vacuum}  }

\def\preprint{\twocolumn\sloppy\flushbottom\parindent 2em
        \leftmargini 2em\leftmarginv .5em\leftmarginvi .5em
        \oddsidemargin -.5in    \evensidemargin -.5in
        \columnsep .4in \footheight 0pt
        \textwidth 10.in        \topmargin  -.4in
        \headheight 12pt \topskip .4in
        \textheight 6.9in \footskip 0pt
        \def\@oddhead{\thepage\hfil\addtocounter{page}{1}\thepage}
        \let\@evenhead\@oddhead \def\@oddfoot{} \def\@evenfoot{} }

\def\numberbysection{\@addtoreset{equation}{section}
        \def\theequation{\thesection.\arabic{equation}}}

\def\underline#1{\relax\ifmmode\@@underline#1\else
        $\@@underline{\hbox{#1}}$\relax\fi}

\def\titlepage{\@restonecolfalse\if@twocolumn\@restonecoltrue\onecolumn
     \else \newpage \fi \thispagestyle{empty}\c@page\z@
        \def\thefootnote{\fnsymbol{footnote}} }

\def\endtitlepage{\if@restonecol\twocolumn \else \newpage \fi
        \def\thefootnote{\arabic{footnote}}
        \setcounter{footnote}{0}}  

\catcode`@=12
\relax

\def\figcap{\section*{Figure Captions\markboth
        {FIGURECAPTIONS}{FIGURECAPTIONS}}\list
        {Figure \arabic{enumi}:\hfill}{\settowidth\labelwidth{Figure
999:}
        \leftmargin\labelwidth
        \advance\leftmargin\labelsep\usecounter{enumi}}}
 \relax
\def\tablecap{\section*{Table Captions\markboth
        {TABLECAPTIONS}{TABLECAPTIONS}}\list
        {Table \arabic{enumi}:\hfill}{\settowidth\labelwidth{Table
999:}
        \leftmargin\labelwidth
        \advance\leftmargin\labelsep\usecounter{enumi}}}
 \relax
\def\reflist{\section*{References\markboth
        {REFLIST}{REFLIST}}\list
        {[\arabic{enumi}]\hfill}{\settowidth\labelwidth{[999]}
        \leftmargin\labelwidth
        \advance\leftmargin\labelsep\usecounter{enumi}}}
 \relax
\makeatletter
\newcounter{pubctr}
\def\publist{\@ifnextchar[{\@publist}{\@@publist}}
\def\@publist[#1]{\list
        {[\arabic{pubctr}]\hfill}{\settowidth\labelwidth{[999]}
        \leftmargin\labelwidth
        \advance\leftmargin\labelsep
        \@nmbrlisttrue\def\@listctr{pubctr}
        \setcounter{pubctr}{#1}\addtocounter{pubctr}{-1}}}
\def\@@publist{\list
        {[\arabic{pubctr}]\hfill}{\settowidth\labelwidth{[999]}
        \leftmargin\labelwidth
        \advance\leftmargin\labelsep
        \@nmbrlisttrue\def\@listctr{pubctr}}}
 \relax
\makeatother
\newskip\humongous \humongous=0pt plus 1000pt minus 1000pt

\newif\ifdtup

\relax

\def\be{\begin{equation}}
\def\ee{\end{equation}}
\def\ba{\begin{eqnarray}}
\def\ea{\end{eqnarray}}

\def\del{\partial}

\def\s{\sigma}

\def\bs{\bigskip}
\def\no{\noindent}

\def\qq{\qquad}

\def\br{\bigr}

\def\IR{\relax{\rm I\kern-.18em R}}

\def \bd {\bar \del}

\def \ha {{1\over 2}}
\def \half {{1\over 2}}

\def \ov {\over}

\def\IR{\relax{\rm I\kern-.18em R}}
\def\IL{\relax{\rm I\kern-.18em L}}

\def\inv{^{\raise.15ex\hbox{${\scriptscriptstyle -}$}\kern-.05em 1}}

\def\Tr{{\rm Tr}}

\begin{document}

\renewcommand{\theequation}{\thesection.\arabic{equation}}
\csname @addtoreset\endcsname{equation}{section}

\newcommand{\beq}{\begin{equation}}
\newcommand{\eeq}[1]{\label{#1}\end{equation}}
\newcommand{\ber}{\begin{equation}}
\newcommand{\eer}[1]{\label{#1}\end{equation}}
\newcommand{\eqn}[1]{(\ref{#1})}
\begin{titlepage}
\begin{center}

\rightline{CCNY-HEP-16/8}
\rightline{September 2016}

${}$
\vskip .2 in

{\large\bf Composition of many spins, random walks and statistics}

\vskip 0.4in

{\bf Alexios P.  Polychronakos$^1$} \ and\ {\bf Konstantinos Sfetsos}$^{2}$
\vskip 0.1in

\vskip 0.1in
{\em
${}^1$Physics Department, the City College of New York, New York, NY 10031, USA\\
\vskip -.04 cm and\\
\vskip -.04 cm
The Graduate Center, CUNY, New York, NY 10016, USA
}
\vskip 0.15in

 {\em
${}^2$Department of Nuclear and Particle Physics,\\
Faculty of Physics, National and Kapodistrian University of Athens,\\
Athens 15784, Greece\\
}

\vskip 0.1in

{\footnotesize \texttt  apolychronakos@ccny.cuny.edu, ksfetsos@phys.uoa.gr}


\vskip .5in
\end{center}

\centerline{\bf Abstract}

\no
The multiplicities of the decomposition of the product of an arbitrary number
$n$ of spin $s$ states into irreducible $SU(2)$ representations are computed. Two complementary methods
are presented, one based on random walks in representation space and another based on the partition function
of the system in the presence of a magnetic field. The large-$n$ scaling limit of these multiplicities is derived,
including nonperturbative corrections, and related to semiclassical features of the system.
A physical application of these results to ferromagnetism is explicitly worked out.
Generalizations involving several types of spins, as well as spin distributions, are also presented.
The corresponding problem for (anti-)symmetric composition of spins is also considered and shown to obey
remarkable duality and bosonization relations and exhibit novel large-$n$ scaling properties.

\vskip .4in
\noindent
\end{titlepage}
\vfill
\eject

\newpage

\tableofcontents

\noindent

\def\baselinestretch{1.2}
\baselineskip 20 pt
\noindent


\setcounter{equation}{0}
\section{Introduction}
The problem of decomposing a direct product of $SU(2)$ irreducible representations (spins) into irreducible
components is ubiquitous in physical and mathematical applications. Situations involving many spins, possibly
coupled with scalar $SU(2)$-invariant interactions arise in condensed matter situations, spin chains,
elasticity theory, nuclear physics, matrix models, and even loop quantum gravity.

The group theory of decomposition of a sum of spins into components is, of course, well established and elementary.
The situation becomes more interesting, however, when we consider a large number of (possibly different) spins
and are interested
in deriving the exact number of irreducible components, as well as their scaling properties for macroscopically
many and possibly indistinguishable spins. This will be the focus of our analysis.

Our motivation for
carrying out this analysis is manifold. First, we wished to present a complete, intuitive and pedagogical treatment
of the problem. In this vein, we chose to work using in parallel two distinct approaches, arising from
different perspectives and offering complementary intuitions: one based directly on representation composition
rules, as familiar to all students of physics and mathematics, and another based on the statistical mechanics of
the system, invoking relevant physical concepts. The first approach turns out to be equivalent to considering
a type of discrete random walk in representation space, which becomes a Brownian motion in the scaling limit,
while the second approach essentially deals with the statistical mechanics of a gas of particles, invoking
thermodynamics in the scaling limit.

In addition, we wanted to derive a comprehensive set of formulae and results,
including exact combinatorial expressions, asymptotic expressions, as well as perturbative and nonperturbative
corrections to asymptotics. Such results, to our knowledge, were not readily available in the literature.

Finally, we wanted to examine various generalizations of this problem of physical and mathematical interest,
with an eye on possible physical applications. We present such an application in which the transition
from a paramagnetic to a ferromagnetic regime and  the corresponding Curie temperature
are explicitly computed for a system of interacting spins.
We also treat in detail the situation where we have many types of spins, or even a large distribution of spin values,
and study its behavior in the scaling limit,
demonstrating that it depends on the properties of the spin distribution and giving criteria for the validity of the
generic Gaussian-like scaling limit. Further, we examine a system of identical spins obeying bosonic or fermionic
statistics, which requires extracting the irreducible components in symmetric or antisymmetric tensor products
of spin representations. We derive useful bosonization and duality relations and study the scaling properties of
such systems in detail. Interestingly, both the number of spins and the size of each spin must
become big to have a meaningful scaling limit, whose properties turn out to be quite different than those
for the simple direct product of (distiguishable) spins.

We present our results in the upcoming sections, hinting at additional
physical applications and possible extensions of our analysis to other groups and situations in the conclusions.

\section {The problem of composing spins}
\renewcommand{\theequation}{\thesection.\arabic{equation}}

Consider a system of $n$ spins $s$ ($s=\half, 1, \frac{3}{2}, \dots$), all of the same value of $s$,
and their decomposition into
irreducible components. Call $d_{s,n,j}$ the number of spin $j$ components in their direct product.
The task is to calculate $d_{s,n,j}$ and find its large-$n$ and large-$j$ asymptotic behavior.

\no
The problem of finding the multiplicities of a product of spins has certainly been examined
before \cite{Kiril,Mend} and very recently in \cite{CuZa}.
Nevertheless,
we will solve it below in a self-contained treatment, presenting two distinct methods of derivation.
This analysis has pedagogical
value, as the two methods are related but intuitively distinct: one builds directly on the known
composition rules of $SU(2)$ representations, while the other is more
physically motivated, working in terms of a partition function.
The two methods are also useful for understanding the large-$n$ limit,
the first one invoking concepts of Brownian motion and the second one concepts
of thermodynamics, and
for attacking the generalizations of the problem considered in the next sections.

\subsection{First derivation method: Direct composition of representations}

From the standard composition rule for spins
\be
[ j_1 ] \otimes [ j_2 ] = \sum_{j=|j_1 - j_2|}^{j_1 + j_2} \oplus ~ [ j ]\ ,
\ee
we deduce the recursive relation for the number of spins $j\geqslant s$
\be
d_{s,n+1,j} = \sum_{\ell=j-s}^{j+s} d_{s,n,\ell}~\ , ~~~j \geqslant s \, .
\label{compo}
\ee
This, essentially, describes a random walk on a two-dimensional lattice, with $n$
playing the role of discrete time and $j$ the role of position,
where a particle on the lattice point ($n,j$) can jump to $2s+1$ possible points ($n+1,j-s$), $\dots$
($n+1,j+s$), and $d_{s,n,j}$ enumerates all the walks that go through the point ($n,j$). We start with a singlet
at $n=0$ (which obviously gives a single spin $s$ at $n=1$, as required), corresponding to the particle starting at the point ($0,0$) on the lattice, and put $d_{s,0,j} = \delta_{j,0}$.

\no
What spoils the above picture is the fact that for $j<s$ the composition rule (\ref{compo}) is modified:
\be
d_{s,n+1,j} = \sum_{\ell=s-j}^{j+s} d_{s,n,\ell}~\ , ~~~ j<s \, .
\label{compp}
\ee
In the random walk picture, this amounts to removing from (\ref{compo}) the jumps to positions $j-s , \dots , s-j-1$
leaving only $2 j+1$ final positions. That is,
particles never dip below $j=0$, but also move {\it away} from zero if they start too close to it.

\no
A simple trick remedies the situation and allows us to use the random walk picture based on \eqn{compo} without modifications.
We simply extend the range of $j$ to include {\it negative} values, and consider configurations
obeying the antisymmetry condition
\be
d_{s,n,j} = - d_{s,n,-j-1} \ ,
\label{asy}
\ee
around $j= - \half$.
It is easy to see that for $j\geqslant s$ the existence of these negative $d_{s,n,-j}$ does not alter the result, since they
do not participate in the sum in (\ref{compo}). For $j<s$, however, their contribution exactly cancels the terms in
(\ref{compo}) that are missing in (\ref{compp}). Therefore, (\ref{compo}) with antisymmetric $d_{s,n,j}$ reproduces
the correct result (\ref{compp}) for $j<s$. Note, further, that the rule (\ref{compo})
preserves the antisymmetry condition (\ref{asy}). Also, due to the latter relation we have that $d_{s,n,-\half} = 0$.

\no
In conclusion, we can consider the unmodified recursion relation (\ref{compo}) but, now, with properly
antisymmetrized initial conditions
\be
d_{s,0,j} = \delta_{j,0} - \delta_{j+1,0}\ .
\ee
In terms of random walks, we have a "positive" walk starting at ($0,0$) and a "negative" walk starting
at ($0,-1$) that proceed independently up and down, and we subtract the number of negative walks that reach the point ($n,j$)
from the number of positive ones reaching the same point.

\no
Since the solution of (\ref{compo}) is linear in the initial conditions for $d_{s,0,j}$, we can find separately the solution
$d^+_{s,n,j}$ for $d_{s,0,j} = \delta_{j,0}$ and $d^-_{s,n,j}$ for $d_{s,0,j} = \delta_{j+1,0}$ and then subtract the results, leading to
\be
d_{s,n,j} = d^+_{s,n,j} - d^-_{s,n,j}\ .
\ee
Due to the translational invariance of (\ref{compo}) in the
$j$ direction, if $d_{s,n,j}$ is a solution, so is $d_{s,n,j+c}$ for any fixed $c$. Since both of the above random walks
start from a single point at $n=0$, with $d_{s,n,j}^-$ starting one unit of $j$ lower that $d_{s,n,j}^+$, the
two are related as
\be
d^-_{s,n,j-1} = d^+_{s,n,j} := D_{s,n,j}\ ,
\ee
where we called $D_{s,n,j}$ the solution of the unrestricted recursion relation (\ref{compo}) with $d_{s,0,j} = \delta_{j,0}$.
We conclude that
\be
d_{s,n,j} = D_{s,n,j} - D_{s,n,j+1}\ .
\label{dDj}
\ee

\no
It remains to evaluate $D_{s,n,j}$. This can be done via a discrete Fourier transform in $j$ of equation
(\ref{compp}) or, equivalently, using generating functions. Define
\be
\zeta_{s,n} (x) = \sum_{j=-\infty}^\infty x^j d_{s,n,j}\ ,\qquad
x\in \mathbb{C}\ .
\label{defgen}
\ee
Hence $x$ plays the role of $e^{ik}$ in the Fourier transform.
Then (\ref{compo}) implies that
\be
\zeta_{s,n+1} (x) = (x^{-s} + \cdots + x^s ) \, \zeta_{s,n} (x) = \frac{x^{s+1} - x^{-s}}{x-1} \, \zeta_{s,n} (x)  \ ,
\ee
which immediately yields
\be
\zeta_{s,n} (x) = \left( \frac{x^{s+1} - x^{-s}}{x-1} \right)^n \zeta_{s,0} (x)   \ .
\label{asz}
\ee
From the initial conditions we find that $\zeta_{s,0} = 1-x^{-1}$, so
\be
\zeta_{s,n} (x) = \frac{\left( x^{s+1} - x^{-s} \right)^n}{x (x-1)^{n-1}}  \ .
\label{gens}
\ee
The task, now, is to perform a Laurent expansion of the above function
in powers of $x$ and calculate the coefficient of $x^j$. Only non-negative $j$ are of interest,
but at any rate the generating function satisfies the relation
\be
\zeta_{s,n} (x^{-1}) = - x \, \zeta_{s,n} (x)\ , \ee
ensuring the proper antisymmetry conditions.

\no
One issue with expanding (\ref{gens}) in powers of $x$ is that there are various strategies that produce
equivalent but, at face value, quite different results. For instance, as the expression in (\ref{gens}) stands we could
expand the denominator in powers of $x$ which, combined with the expansion of the power in the numerator,
would give an infinite series. Yet we know that the highest power on $x$ that would appear is $x^{ns}$, which
means that the coefficients of terms higher than that must vanish identically.
To obtain an expression that explicitly terminates at power $x^{ns}$ we write $\zeta_{s,n} (x)$ instead as
\be
\zeta_{s,n} (x) = \frac{\left( x^{s+1} - x^{-s} \right)^n}{x^n (1-x^{-1})^{n-1}}
\label{genss}
\ee
and expand the denominator in powers of $x^{-1}$. This will obviously produce a highest power of $x^{ns}$.
It will also give an infinite series in negative powers, that must similarly terminate due to (\ref{asz}).
Going through the steps and isolating the coefficient of $x^j$ we obtain that
\be
\boxed{
d_{s,n,j} = n(n-1) \sum_{k=0}^{\left[\frac{sn-j}{2s+1}\right]}
(-1)^k \, \frac{\Bigl( (s+1)n - (2s+1)k -j-2 \Bigr)!}{k! (n-k)! \left(sn - (2s+1)k -j \right)!}
} \ ,
\label{opa}
\ee
where we recall that $[r]$ denotes the integer part of the real number $r$.
This is the most compact formula we could find for the multiplicity coefficients.

\no
Our formula is equivalent but not identical in form to the result of \cite{Mend}
reproduced in \cite{CuZa}.
As we already explained, however, this expression is not unique.  For small, specific values of $s$ alternative formulae can be produced by working directly with $(x^s + \cdots + x^{-s} )^n$
or using other tricks. As an example, consider the case with $s=1$. We register three alternative formulae. The first one is
\be
d_{1,n,j} = n(n-1) \sum_{k=0}^{\left[\frac{n-j}{3}\right]}
(-1)^k \, \frac{(2n - 3k -j-2)!}{k! (n-k)! \left(n - 3k -j \right)!}\ ,
\ee
which is simply (\ref{opa}) evaluated for $s=1$.
A second expression can be obtained by writing $\zeta_{1,n}(x)=(1-x^{-1}) (1+x+x^{-1} )^n$ and then
using the identity
\be
(x + 1 + x^{-1} )^n = \sum_{k,m=0}^{k+m \leqslant n} \frac{n!}{k! m! (n-k-m)!} \, x^{k-m}\ .
\label{inddd}
\ee
The result is
\be
d_{1,n,j} = n! \sum_{k=0}^{\left[\frac{n-j}{2}\right]}
\frac{3k+2j+1-n}{k! (k+j+1)! (n-2k-j)!}\ .
\ee
A third expression can be obtained by using instead of \eqn{inddd} the identity
\ba
(x + 1 + x^{-1} )^n &=& \left[ \left( x^\half + x^{-\half} \right)^2 -1 \right]^n
= \sum_{k=0}^n (-1)^k  \frac{n!}{k! (n-k)!} \left( x^\half + x^{-\half} \right)^{2k}
\nonumber \\
&=&  \sum_{m,k=0}^n (-1)^k  \frac{n! (2k)!}{k! (n-k)! m! (2k-m)!} x^{k-m}\ ,
\ea
resulting to the expression
\be
d_{1,n,j} = (-1)^n n! \sum_{k=j}^{n} (-1)^k
\frac{(2k)! (2j+1)}{k! (n-k)! (k-j)! (k+j+1)!}\ .
\ee
This is equivalent to treating the $s=1$ spin as one of the components of the composition of
two $s=\half$ spins.

\no
We also record the result for $s=\half$, which is particularly simple
\be
d_{\half,n,j} = \frac{n! (2j+1)}{\left(\frac{n}{2}-j \right)!\left(\frac{n}{2}+j+1 \right)!} \ .
\label{dhalf}
\ee

We conclude by presenting two results, immediately obtainable from our analysis.

\no
The full generating function, for both $j$ and $n$, is
\be
{\cal F}_s(x,y) = \sum_{n=0}^\infty \sum_{j=-sn-1}^{sn} d_{s,n,j} \, x^j \, y^n = \frac{x^{s-1}(x-1)^2}{x^s (x-1)-y(x^{2s+1} -1)}\ .
\ee
This owes its simple, explicit form to the fact that it includes both positive and negative $j$. The generating function
over non-negative $j$, i.e., the analytic part of ${\cal F}_s(x,y)$, can be expressed in terms of the Hilbert transform
of $\cal F$ in $x$. This would be an unnecessary complication, at any rate, since both functions recover the coefficients
for non-negative $j$ in the standard way
\be
d_{s,n,j} = \frac{1}{4\pi^2} \int_0^{2\pi} d\varphi \int_0^{2\pi} d\phi \, {\cal F}_s (e^{i\varphi},e^{i\phi}) \,
e^{-ij \varphi-in \phi}\ .
\ee

\no
A generalized number of states relation is obtained directly from the generating function $\zeta_{s,n} (x)$.
By expressing the sum over negative $j$ in terms of $-d_{s,n,-1-j}$, writing $q$ for $x$ (purely for aesthetic reasons,
and to evoke $q$-deformations) and using the definition of $q$-deformed numbers
\be
[ N ]_q = q^{-\frac{N-1}{2} }+ q^{-\frac{N-3}{2}} + \cdots + q^{\frac{N-3}{2}} +
q^{\frac{N-1}{2}} = \frac{q^{\frac{N}{2}} - q^{-\frac{N}{2}}}{q^{\frac{1}{2}} - q^{-\frac{1}{2}}}\ ,
\label{qdef}
\ee
we obtain the identity
\be
\sum_{j=0}^{sn} [ 2j+1 ]_q \, d_{s,n,j} = \left( [2s+1]_q \right)^n\ ,
\label{qrel}
\ee
which is valid for any $q$.
For $q=1$ we recover the total number of states relation
\be
\sum_{j=0}^{sn} (2j+1) \, d_{s,n,j} = (2s+1)^n\ .
\ee
Other special values of $q$ give other special relations. For $q=-1$, in particular, we get the interesting
parity relation
\ba
\sum_{j=0}^{sn} \, (-1)^j d_{s,n,j} &=& (-1)^{sn} \ , \qq  s={\rm integer}\ ,
\cr
&=& 0 \ ,\qq\qq\phantom{x}  s={\rm {half{-}integer}}, ~n={\rm even}\ .
\ea
Higher moments of the distribution $d_{s,n,j}$ can be found by differentiating \eqn{qrel} with respect
to $q$ and then setting $q=1$. Odd derivatives vanish,
but even ones give nontrivial results. For example, the second derivative gives the relation
\be
\sum_{j=0}^{sn} \frac{j(j+1)(2j+1)}{3} \, d_{s,n,j} = n\frac{s(s+1)}{3} \, (2s+1)^n\ ,
\ee
which expresses the fact that for $n$ independent spins $s_a$
\be
\Tr \left( \sum_{a=1}^n s_a \right)^2 = \sum_{a=1}^n \Tr \, s_a^2\ .
\ee

\subsection{Second derivation method: Partition function}

We introduce a magnetic field $B$ in the $z$-direction coupled to each spin with energy $-B s_3$ and calculate
the partition function of the system with temperature parameter $\beta$. The total energy of the system is
\be
E = -B \sum_{a=1}^n s_{a,3} = -B m\ ,
\ee
where $s_{a,3}$ is the $z$-component of spin $a$, and we called $m$ the total $z$ component of the system.
Call $D_{s,n,m}$ the number of states with total $z$-component $m$. Then the partition function is
\be
Z_{s,n} = \sum_{m=-ns}^{ns} \, D_{s,n,m} e^{-\beta E_m} = \sum_{m=-ns}^{ns} \, D_{s,n,m} e^{\beta B m}
= \sum_{m=-ns}^{ns} \, D_{s,n,m} x^m ~,~~~ x := e^{\beta B}\ .
\ee
On the other hand, the system consists of $n$ non-interacting, distinguishable spins $s$, so the total
partition function is simply give by
\be
Z_{s,n} = Z_{s,1}^n = \left( \sum_{m=-s}^s e^{\beta Bm} \right)^n = (x^s + \cdots + x^{-s} )^n
= \left( \frac{x^{s+1} - x^{-s}}{x-1} \right)^n\ .
\ee
Altogether, we obtain
\be
Z_{s,n} (x) = \sum_{m=-ns}^{ns} \, D_{s,n,m} x^m =  \frac{(x^{s+1} - x^{-s})^n}{(x-1)^n}\ ,
\label{Zn}
\ee
which is the generating function for the number of states with total $z$-component equal to $m$.

\no
Next, recall that, every spin-$j$ in the decomposition of the $n$ spins
will contribute exactly one state at level $m$,
for every $|m| \leqslant j$. So the decrease in number of states
from $m=j$ to $m=j+1$ is exactly the number of spins-$j$
(which contribute to $m=j$ but not to $m=j+1$). We conclude that
\be
d_{s,n,j} = D_{s,n,j} - D_{s,n,j+1}\ .
\ee
The above is valid for $j \geqslant 0$, but we can formally extend
it to $j<0$, in which case, since obviously $D_{s,n,m} = D_{s,n,-m}$,
it gives $d_{s,n,j} = -d_{s,n,-j-1}$.

\no
The generating function for the number of spins $j$, then,
is as in \eqn{defgen}
\ba
\zeta_{s,n} (x) & = & \sum_{j=-\infty}^\infty x^j d_{s,n,j}
=  \sum_{j=-\infty}^\infty x^j
\left( D_{s,n,J} - D_{s,n,j+1} \right)
= (1-x^{-1} ) Z_{s,n} (x)
\nonumber\\
& = &   \frac{(x^{s+1} - x^{-s})^n}{x(x-1)^{n-1}}\ ,
\ea
thus recovering the result we found with the first method
based on random walks.
Note, further, that the relation (\ref{qrel}) that
we found before is nothing but the total partition function of the system,
with $q=e^{\beta B}$.

\no
The partition function method makes the formulae
we found before more intuitive, assigning them a
specific physical significance, and will also be useful
when we consider modifications of the original system, as we will see below.

\section{Asymptotics for large $n$}

It is interesting to consider the asymptotic behavior of $d_{s,n,j}$ for large $n$.
Before performing the calculation based on our previous exact results, however, we will exploit the fact that
a system of a large number of spins becomes essentially classical to present a semiclassical argument that is
more intuitive, highlights the classical features of this limit and recovers the exact result.

\subsection{A semiclassical argument}

In the large-$n$ limit, the total angular momentum
of the system becomes large, and thus behaves like a classical spin vector $\vec S$ whose
components can be treated as
classical commuting variables. This, however, does not mean that the length of $\vec S$
tends to a unique
value. Instead, when considering the full ensemble of quantum states for each spin,
the total angular momentum
will be a {\it statistical distribution} of classical spins $\vec S$ of various
sizes and orientations.

To derive this distribution, we argue in analogy to the distribution of velocities
of a gas of free classical
particles. The Cartesian components of the angular momentum become independent variables that,
by the law of large numbers, are normally distributed. We need only one input from the
microscopic properties of individual spins, namely that, considering their $2s+1$
individual quantum states, any specific
component of each spin (say, $s_z$) takes values from $+s$ to $-s$ with equal weights, and thus
has zero mean and variance
\be
\sigma^2 = \frac{1}{2s+1} \sum_{m=-s}^s m^2 = \frac{s(s+1)}{3}\ .
\label{sig}
\ee
Since the spins are uncorrelated, the Central Limit Theorem implies that the corresponding component of
the total angular momentum will be canonically distributed with zero mean and variance $n \sigma^2$
\be
f(S_z ) = \frac{1}{\sigma \sqrt{2\pi n}} \, e^{-\frac{S_z^2}{2n\sigma^2}}\ ,
\ee
which in fact holds for any spin component. The distribution for all three components of the angular momentum is
simply given by the product
\be
f(S_x , S_y , S_x ) = \frac{1}{\sigma^3 (2\pi n)^{3/2}} \, e^{-\frac{S_x^2 +S_y^2 +S_z^2}{2n\sigma^2}}\ .
\ee
The above implies a Maxwell distribution for the length of $\vec S$. Indeed, by writing $dS_x dS_y dS_z = S^2 dS d\Omega$
and integrating over angles we obtain that
\be
f(S) = \frac{4\pi S^2}{\sigma^3 (2\pi n)^{3/2}} e^{-\frac{S^2}{2n\sigma^2}}\ .
\label{f}
\ee
It remains to make contact with the number of quantum spins-$j$. We note, first, that the length of the
classical total angular momentum vector corresponding to $j$ is, for large $j$, $S=j$, so we do this replacement
in (\ref{f}).
Further, we are interested in the total number of spins, rather than their distribution. So we need to multiply
(\ref{f}) by the total number of quantum states $(2s+1)^n$. Finally, we recall that each spin $j$
corresponds to $2j+1$ states, and the above distribution takes them all into account. So we need
to divide (\ref{f}) by $2j+1 \sim 2j$ (for large $j$) to count each spin once. Altogether, and using
the expression (\ref{sig}) for $\sigma$, we obtain that
\be
d_{s,n,j} = \frac{(2s+1)^n}{2j} f(j) =  \left(\frac{3}{s(s+1)n} \right)^\frac{3}{2} \frac{(2s+1)^n}{\sqrt{2\pi}}
\, j \, e^{-\frac{3 j^2}{2s(s+1) n}}\ .
\label{sc}
\ee
As we shall see, this is in fact the exact result in the large-$n$, large-$j$ limit.

\subsection{The exact derivation}

In the large-$n$ limit, large values of $j$ will dominate the decomposition. So $j$ becomes essentially a
continuous variable. Calling $D_{s,n} (j)$ and $d_{s,n} (j)$ the corresponding combinatorial quantities in that
limit, the two are related as
\be
d_{s,n} (j) = D_{s,n,j} - D_{s,n,j+1} \simeq - \frac{\partial}{\partial j}
D_{s,n} (j+{\textstyle \half})\ ,
\label{dD}\ee
where we Taylor-expanded $D_{s,n} (j)$ around the midpoint of the difference $j+{\textstyle \half}$ and
kept the leading term only. We note that
the distinction between $j$ and $j+{\textstyle \half}$ is immaterial in the large-$n$, large-$j$ limit, but this
choice gives the most accurate approximation of the difference in terms of the derivative.

\no
We could, in principle, calculate the large-$n$ limit of $d_{s,n,j}$ starting directly
from the exact combinatorial formula (\ref{opa}) and using the Stirling approximation
for the factorials appearing in it. Due to the existence of sums with alternating signs,
however, obtaining the correct limit is somewhat nontrivial and requires a careful analysis.
Instead, it is advantageous to work directly with the random walk approach or the partition function.
We will present both methods as they each present advantages it terms of intuition.

\vskip 0.4cm

\subsubsection{Random walk approach}

The result for $D_{s,n} (j)$  can be obtained quite easily in the random walk picture.
In the large-$n$ limit the walk becomes essentially a Brownian motion, that is, the sum of many independent
identically distributed individual random steps (one for each step $n \to n+1$).
Each random step, going from $+s$ to $-s$ with equal weights, has zero mean and variance given by \eqn{sig}.
The total random walk $D_{s,n,j}$ (starting from $j=0$ at $n=0$), then, has mean zero and variance $\s_n^2 = n \s$.
By the Central Limit Theorem, the total Brownian walk becomes normally distributed with zero mean and
variance as above.
The total normalization is proportional to the total number of states $(2s+1)^n$ as it must
satisfy the number of states relation
\be
\sum_{j=-sn}^{sn} D_{s,n,j} \simeq  \int_{-\infty}^\infty \, D_{s,n}(j) \, dj = (2s+1)^n\ .
\ee
Hence, for large $n$ and $j$
\be
D_{s,n,j}  \simeq D_{s,n}(j) = \sqrt{\frac{3}{2\pi ns(s+1)}} \, (2s+1)^n\
\, e^{-\frac{3j^2}{2s(s+1) n}}
\label{wD}
\ee
Upon taking the derivative as instructed by (\ref{dD}) we obtain that
\be
\boxed{
d_{s,n,j} \simeq d_{s,n}(j) = \left(\frac{3}{s(s+1)n} \right)^\frac{3}{2}
\frac{(2s+1)^n}{\sqrt{2\pi}}
\, (j+{\textstyle \half}) \, e^{-\frac{3(j+{\half})^2}{2s(s+1) n}}
}\ .
\label{nasy}
\ee
We observe that, upon dropping the subleading term $\half$ in $j$, this is the same result (\ref{sc})
obtained in the semiclassical calculation. This result was also obtained in \cite{CuZa}.

It should be noted that, for integer $ns$, $j$ will take only integer values, while for half-integer
$s$ and odd $n$, $j$ will take only half-integer values. The spacing between successive values of $j$, however,
is 1 in all cases, so the above formula expresses accurately the number of irreps $j$ for the appropriate values
of $j$ in the large-$n$ limit.
As a check, we can see that the above satisfies the total number of states condition
\be
\sum_{j=0}^{sn} (2j+1) d_{s,n,j} \simeq  \int_{-\half}^\infty (2j+1) \, d_{s,n}(j) \, dj = (2s+1)^n\ .
\ee
The value of $j=J_{s,n}$ for which $d_{s,n,j}$ becomes maximum
for fixed $s$ and $n$ in that limit is given by
\be
J_{s,n} = \sqrt{\frac{s(s+1) n}{3}}\ ,
\label{jsn}
\ee
where we have dropped the distinction between $j$ and $j+\ha $). The corresponding maximum number of states is
\be
d_{s,n,{\rm max}} = d_{s,n,J_{s,n}} = \frac{3 (2s+1)^{n}}{s(s+1) n\sqrt{2 \pi e}}\ .
\ee
However, the distribution
of $j$ is {\it not} sharply peaked around $J_{s,n}$, since its mean and standard deviation are both of order
$J_{s,n}$. Specifically,
\ba
&& \langle j \rangle = \frac{\sum_j j d_{n,j}}{\sum_j d_{n,j}} = \sqrt{\frac{s(s+1) n \pi}{6}} =
\sqrt{\frac{\pi}{2}} J_{s,n} \ ,
\nonumber\\
&& \langle j^2 \rangle = \frac{\sum_j j^2 d_{n,j}}{\sum_j d_{n,j}} =
2{\frac{s(s+1) n}{3}} = 2 J_{s,n}^2
\ea
and thus
\be
\Delta J = \sqrt{\langle j^2 \rangle - \langle j \rangle^2} = \sqrt{\frac{4-\pi}{2}} J_{s,n}\ .
\label{devv}
\ee
So the spin system is not dominated by a single classical spin component but remains a distribution over
various spins, most of them classical since $j \sim J_{n,s} \sim \sqrt{n}$, as in the semiclassical argument.

\subsubsection{Partition function approach}

The partition function $Z_{s,n} (x)$ is the generating function for
the number of states with total $z$-component equal to $m$. From it we can recover the number of states as
\be
D_{s,n,m} = \frac{1}{2\pi} \int_0^{2\pi} d\phi \, Z_{s,n} (e^{i\phi}) \, e^{-im\phi}\ .
\label{Fou}
\ee
The expression of the full partition function in (\ref{Zn}) is the $n$-th power of a function $Z_{s,1}(x)$. So
in the large-$n$ limit we can use the saddle point approximation for the above integral.

This is, in fact, related to determining the entropy of the system as a function of its energy (which is proportional
to $m$) in the thermodynamic limit. To make the analogy explicit, we set $x=e^{\beta}$, with $\beta$ a redefined temperature parameter dual to $-m$ (that is, we set the magnetic field $B=1$), the sign chosen such that positive
values of $\beta$ correspond to positive $m$. We define the single-spin free energy
\be
-\beta F_{s} (\beta) =  \ln Z_{s,1} (e^{\beta}) = \ln \frac{e^{(s+1)\beta} - e^{-s\beta}}{e^{\beta}-1}
= \ln {\sinh (2s+1){\beta\ov 2} \ov \sinh {\beta\ov 2}}\ ,
\ee
while the total free energy of the system is
\be
F_{s,n} (\beta ) = n F_s (\beta )\ .
\ee
The entropy $S$ is given by the standard relation, with $-m$ playing the role of $E$
\be
S = - \beta F_{s,n} -\beta m=  - n \beta F_s -\beta m
\label{SmF}
\ee
and corresponds to the logarithm of the number of states. The thermodynamic equilibrium for fixed $m$ is
achieved by maximizing $S$ in terms of $\beta$, which leads to the thermodynamic relation
\be
\frac{\partial S}{\partial \beta} = 0 ~~~\Longrightarrow ~~~
\frac{m}{n} = -\frac{\partial (\beta F_s )}{\partial \beta}\ .
\label{SF}
\ee
For large values of $n$, the solution to the above will be for a small value of $\beta$. Performing
a Taylor expansion of $\beta F_s$ around $\beta =0$ we have
\be
-\beta F_s = \ln (2s+1) + \frac{1}{6} s(s+1) \beta^2 + \dots\ .
\ee
So (\ref{SF}) gives $\beta$ in the large-$n$ limit as
\be
\beta \simeq  \frac{3m}{ns(s+1)}\ .
\ee
Putting this value back into the expression for the entropy we obtain
\be
S \simeq n \ln (2s+1) - \frac{3 m^2}{2ns(s+1)}
\ee
and the corresponding number of states
\be
D_{s,n,m} = e^S \simeq (2s+1)^n \, e^{-\frac{3m^2}{2ns(s+1)}}\ .
\label{preD}
\ee

\no
It should be appreciated that the above procedure reproduces the steps in the saddle-point
evaluation of the Fourier integral in (\ref{Fou}), putting them into a statistical mechanics context.
The one thing missed in the thermodynamic
approach is the determinant arising from the quadratic integration, which would contribute
a logarithmic term in the entropy (a prefactor in the number of states) that is
thermodynamically subleading. This determinant ensures that the large-$n$ formula
reproduces the correct number of states and can be read off from the exponent of the
Gaussian in $m$. Overall, we obtain precisely
the expression in (\ref{wD}). The remaining calculations proceed as
in the random walk approach.

\subsection{$1/n$ and nonperturbative corrections}

It should be clear that the saddle-point approximation is essentially equivalent
to the Central Limit Theorem, leading to a normal distribution. The advantage of
the partition/generating function method is that it can also produce subleading corrections
in $1/n$, by keeping additional terms in the expansion of the free energy, as well as nonperturbative
corrections of order $e^{-n}$.

To obtain the next subleading perturbative correction in $1/n$, we work with the
Fourier integral (\ref{Fou}) (in which $\beta = it$) and expand the free energy to the next
(quartic) order in $t$.
we have
\ba
&& \ln Z_{s,n}  (e^{it} ) = n \ln \frac{\sin (2s+1)\frac{ t}{2}}{\sin \frac{t}{2}}
\nonumber\\
&&
\phantom{xx}
= n\ln (2s+1) -n\frac{s(s+1)}{6} t^2 - n\frac{s(s+1)[2s(s+1)+1]}{360} t^4 + \cdots\ .
\ea
Inserting the above expression in the integral and evaluating it to leading order in the
quartic term we obtain, after some algebra (and re-exponentiating subleading terms)
\beqs
D_{s,n} (m) & = &\sqrt{\frac{3}{2\pi ns(s+1)}} \, (2s+1)^n \,
\exp\left\{-\frac{2s(s+1)+1}{40s(s+1)n}\right\}
\times \nonumber\\
&& \exp\left\{-\left[1-\frac{6s(s+1)+3}{10s(s+1)n}\right] \frac{3m^2}{2s(s+1)n}
- \frac{9[2s(s+1)+1] m^4}{40 s^3 (s+1)^3 n^3}\right\} ~~~~~
\label{D1n}
\eeqs
This amounts to a $1/n$ correction of the normalization and of
the coefficient of the quadratic term, plus a new term quartic
in $m$. Since $m^2 \sim n$, we see that the new term is also of order $1/n$.
The distribution $d_{s,n} (j)$ in this case is calculated as
\be
d_{s,n} (j) = D_{s,n} (j) - D_{s,n} (j+1) \simeq
- \frac{\partial}{\partial j} D_{s,n} (j+{\textstyle \half}) - \frac{1}{24} \frac{\partial^3}
{\partial j^3} D_{s,n} (j+{\textstyle \half})
\ee
We note two facts: first, in this case it is important to keep the $\half$ in the argument: since
$j$ is of order $\sqrt n$, it amounts to a $1/\sqrt n$ correction and must be retained. Second, a higher
approximation of the difference is required, involvind the third derivative of $D_{s,n} (j)$, as this
contributes a correction of order $1/n$ and must also be retained.

Using the expression (\ref{D1n}) we can find $d_{s,n} (j)$
to order $1/n$, and can also calculate $1/n$ corrections for the spin $J_{s,n}$ that maximizes it.
Further, by keeping more terms in the expansion of the free energy in $t$ and evaluating the Fourier
integral to the appropriate order, we can calculate higher order corrections in $1/n$. We will not belabor
these any further, leaving them as an exercise for the reader.

We can also obtain nonperturbative in $n$ corrections, relevant for large deviations from the
maximal multiplicity spin $J_{s,n}$, by returning to the full saddle-point evaluation of the number
of states $D_{s,n,m}$. In fact, our "calculation" here will simply amount to codifying the thermodynamic
result in a compact form. We write, for brevity of expression,
\be
f_s (\beta) = - \beta F_s (\beta) = \ln {\sinh (2s+1){\beta\ov 2} \ov \sinh {\beta\ov 2}}\ .
\label{ssp}
\ee
Then (\ref{SmF}) and (\ref{SF}) become
\be
 \partial_\beta f_s (\beta)= \frac{m}{n}\ ,\qq
S (m) = n  f_s (\beta) - m \beta\ .
\label{lege}
\ee
We see that $f_s (\beta)$ and $S (m)$ are related through a Legendre transform.
That is, we need to solve the first equation for $\beta (m)$ and insert in the second equation
to find $S = \ln D_{s,n,m}$ as a function of $m$. 

If we assume that $m$ is of order $n$ (rather that $\sqrt{n}$, as before), $\frac{m}{n}$ is of
order $1$, and so is $\beta (m)$. Therefore, the entropy $S$ is of order $n$, and the number of states
$D = e^S$ is exponential in $n$. Since the total number of states is $N = e^{S (\beta = 0)} =
e^{S (m=0)}$, $D$ is down by a factor $e^{-[S (0) - S(m)] n}$ with respect to the total
number of states, which is clearly a nonperturbative correction. In essence, the full thermodynamic
approach is an "instanton" approach, working around a "classical" saddle point background.

What is missed in the above is, again, the determinant of the $t= i\beta$ integration, which is thermodynamically
subleading. Expanding the exponent of the integrand, $n f_s (\beta) + \beta m$,  to quadratic order in $\beta$
around the thermodynamic value $\beta (m)$, the relevant measure that must multiply the result is
\be
 J (\beta (m) )= \left[2\pi n \, \partial_\beta^2 f_s (\beta ) \right]^{-\half}\ .
\ee
This can be calculated directly in
terms of $S(m)$ by a standard set of steps. From (\ref{lege}) we have that
\be
\partial_\beta^2 f_s = \frac{1}{n} \partial_\beta m\ .
\label{dS}
\ee
On the other hand, again using (\ref{lege}),
\be
\partial_m S = n \, \partial_\beta f_s \, \partial_m \beta-  \beta -  m \, \partial_m \beta = -\beta
\ee
and, upon differentiating again and using (\ref{dS}), we obtain
\be
\partial_m^2 S = -\partial_m \beta =  -( \partial_\beta m )^{-1} = - (n \, \partial_\beta^2 f_s )^{-1} \ .
\ee
So the full result for $D_{s,n} (m)$ is
\be
D_{s,n,m} = \sqrt{-\frac{\partial_m^2 S(m)}{2\pi}} \, e^{S (m)}\ ,
\label{Dnonpert}
\ee
with $S(m)$ given by the Legendre transform in (\ref{lege}). For generic $s$, this transform cannot
be carried out explicitly, but it can easily be performed numerically since $f_s (\beta )$ and
$\partial_\beta f_s$ are well-behaved functions.

It is instructive and useful to carry out this procedure for two special cases where it can be perfomed
analytically, namely $s=\half$ and $s=1$.

For $s = \half$ we have
\be
f_\half = \ln \left(2 \cosh \frac{\beta}{2} \right) ~,~~~ \partial_\beta f_\half = \half \tanh\frac{\beta}{2}\ ,
\ee
leading to $\beta = \ln(n+2m) - \ln (n-2m)$ and, eventually,
\be
S = n \ln n - \left({n \ov 2} -m \right) \ln \left({n \ov 2} -m \right) - \left({n \ov 2} +m \right)
\ln \left({n \ov 2} +m \right)
\ee
and
\be
- \partial_m^2 S = \frac{n}{ \left({n \ov 2} -m \right)  \left({n \ov 2} +m \right) } \, .
\ee
Altogether, we have that
\be
D_{\half,n} (m) = \frac{n^{n-\half}}{\sqrt{2\pi}\left({n \ov 2} -m \right)^{ {n \ov 2} -m -\half}
\left({n \ov 2} +m \right)^{{n \ov 2} +m -\half}}\ .
\label{half}
\ee
We recognize this as the Stirling approximation (including square-root corrections) of the
combinatorial factor
\be
D_{\half,n,m} = \frac{n!}{\left({n \ov 2} -m \right)! \left({n \ov 2} +m \right)!}\ ,
\ee
which is the exact result (compare with (\ref{dhalf})). The saddle point calculation captures the leading $O(n)$
result plus logarithmic and constant corrections in the exponent (that is, $O(1/{\sqrt n})$ and
$O(1/n)$ prefactors).

The situation for $s=1$ is more interesting. In this case we have
\be
f_1 = \ln (1+2\cosh \beta) ~,~~~ \partial_\beta f_1 = \frac{2 \sinh \beta}{1+2\cosh \beta}\ .
\ee
The equation for $e^\beta$ now becomes quadratic and, picking the relevant (positive) solution,
we obtain that
\be
\beta = \ln \frac{m+ \sqrt{4n^2 - 3m^2}}{2(n-m)}
\ee
and eventually
\be
S =n \ln n + n \ln \frac{n + \sqrt{4n^2 - 3m^2}}{n^2 - m^2} - m \ln \frac{m+\sqrt{4n^2 - 3m^2}}{2(n-m)}
\ee
and
\be
- \partial_m^2 S = \left( 1+\frac{n}{\sqrt{4n^2 - m^2}} \right) \frac{n}{n^2 - m^2}\ .
\ee
As a check, we see that $S(0) = n \ln 3 $ reproducing the correct number of states $e^S = 3^n$.
Also, we can verify that $S(m) = S(-m)$, as expected.
Altogether, we have
\be
D_{1,n} (m) = \frac{1}{\sqrt{2 \pi \sqrt{4n^2 - 3m^2}}}
\left( \frac{3n}{\sqrt{4n^2 - 3m^2}-n}\right)^{n+\half}
\left(\frac{\sqrt{4n^2 - 3m^2}- m}{2(n+m)} \right)^m\ .
\label{one}
\ee
We do not recognize the above as the Stirling approximation of a simple combinatorial formula,
but it nevertheless provides a nonperturbative approximation to the correct result for any $m$.

We conclude by pointing out that formulae (\ref{half}) and (\ref{one}), when expanded to order $m^2$
around $m=0$, reproduce the leading scaling result (\ref{wD}) for $s=\half, 1$, including the correct
prefactor. Further, if we keep terms of order $m^4$ in the expansion, they reproduce the
corresponding results (\ref{D1n}) with $1/n$ corrections, missing only the $m$-independent
correction in the prefactor. This is not true, however, for higher orders of approximation ($m^6$
and beyond), where the saddle point formulae miss several subleading corrections.

\section{A physical application: ferromagnetism}

As a concrete physical application of our results, and in particular of the large-$n$ asymptotics, we present
the derivation of the ferromagnetic transition and properties of a large collection of spins coupled
with spin-aligning mutual interactions \cite{Heis,Rev}.

\no
The Hamiltonian of a system of ferromagnetically coupled spins has the form
\be
H = - \sum_{a \neq b} C_{ab} ~ {\vec s}_a \cdot {\vec s}_b - B \sum_a s_{a,z} \ ,\qq a,b=1,\dots n\ .
\ee
In addition to the magnetic field ${\vec B} = B {\hat z}$ there are $SU(2)$-invariant couplings between spins
that tend to align them, so the coupling constants $C_{ab}$ will be taken positive. In general, only
neighboring spins interact appreciably, so $C_{ab}$ will fall off to zero for spins $a$ and $b$ too far
from each other. We will assume translation invariance, and put
$C_{ab} = C ({\vec a} - {\vec b})$, in a notation where $\vec a$
denotes the position of spin $a$ in some $d$-dimensional lattice configuration. The function $C ({\vec a})$
falls off to zero as the magnitude of $\vec a$ increases.

\no
We will consider the simplified version in which the Hamiltonian assumes the form
\be
H = - \sum_{a \neq b} \frac{c}{2n} ~ {\vec s}_a \cdot {\vec s}_b - B \sum_a s_{a,z} \ ,\qq a,b=1,\dots n\ ,
\label{alleq}
\ee
that is, all couplings are equal and of order $1/n$. The single coupling constant $c$ can be thought
of as the
average weighted coupling of each spin with its neighbors
\be
\sum_{a\neq b} C ({\vec a} - {\vec b}) \equiv
\sum_{a\neq b} \frac{c}{2n} ~~~~~~~{\rm or}~~~~~~~
n \sum_b C ({\vec b} ) = n(n-1) \frac{c}{2n} \simeq n \frac{c}{2}\ .
\ee
This identifies $c$ as
\be
c = 2 \sum_b C ({\vec b})
\ee
and justifies the $1/n$ scaling of the coupling constant in (\ref{alleq}) as the condition ensuring a properly
extensive energy.

\no
The interaction energy corresponding to the above Hamiltonian clearly depends only on the total spin of the system, since
\be
\sum_{a \neq b} {\vec s}_a \cdot {\vec s}_b = \left( \sum_a {\vec s}_a \right)^2 - n s (s+1)
= j(j+1) - n s(s+1)\ .
\ee
Hence, the partition function of the above system is (omitting the trivial constant $ns(s+1)$ in the energy)
\be
Z  = \sum_{j=0}^{ns} \sum_{m=-j}^j  d_{s,n,j} ~e^{\beta \frac{c}{2n} j(j+1) + \beta B m} =
\sum_{j=0}^{ns} d_{s,n,j} \, \frac{\sinh (j+\half)\beta B}{\sinh \frac{\beta B}{2}} \,
e^{\beta \frac{c}{2n} j(j+1)}\ .
\ee
We notice that the summand is symmetric under $j \to -j-1$ (due to the antisymmetry property of
$d_{s,n,j}$), so the sum can be written as
\be
Z  = \frac{1}{2\sinh \frac{\beta B}{2}} \sum_{j=-ns-1}^{ns} d_{s,n,j}  ~
 e^{(j+\half)\beta B + \beta \frac{c}{2n} j(j+1)}\ .
\ee
In the limit $n \to \infty$ we can treat $j$ as a continuous variable and use the asymptotic substitutions
$j+\half \simeq j$, $j(j+1) \simeq j^2$ and $d_{s,n,j} \simeq d_{s,n} (j)$. Then, the partition function
assumes the integral form
\be
Z = \frac{1}{2\sinh \frac{\beta B}{2}} \int_{-\infty}^\infty dj ~ d_{s,n} (j)  ~
 e^{\beta B \, j + \beta \frac{c}{2n} j^2}\ .
\label{Zlargen}
\ee

\subsection{Paramagnetic regime}

Using the asymptotic result (\ref{nasy}) for $d_{s,n} (j)$ in (\ref{Zlargen}) we obtain that
\be
Z = \frac{N}{\sinh\frac{\beta B}{2}} \int_{-\infty}^\infty dj ~j ~ e^{\beta B j +
\frac{1}{2n} \left(\beta c -\frac{3}{s(s+1) } \right) j^2}\ ,
\label{Zferpert}
\ee
where we collected all $\beta$- and $B$-independent constants into the overall factor $N$.

\no
We already observe the signature of a phase transition: the convergence of the integral in (\ref{Zferpert})
is governed by the quadratic exponential. For small $\beta$ (large temperature) it is a converging Gaussian.
For larger $\beta$, however, it becomes divergent, signalling that the system develops sponteaneous magnetization.
The critical value of
$\beta$ where the coefficient of $j^2$ vanishes determines the Curie temperature of the system
\be
T_c = \frac{1}{\beta_c} = \frac{s(s+1)}{3} c\ .
\label{Cur}
\ee
For temperatures above the Curie temperature the partition function can be calculated by the perturbative
integral (\ref{Zferpert}), as
\be
Z = N'\, \frac{\beta B/2}{\sinh\frac{\beta B}{2} } \left(1-{T_c\ov T}\right)^{-\frac{3}{2}}
~ e^{ \frac{n s(s+1)} {6T(T-T_c)}B^2}\ ,
\ee
where the new overall constant $N'$ includes all $T$- and $B$-independent numerical factors.
The magnetization $M = \frac{1}{n}\langle m \rangle$ is calculated in the large-$n$ limit as
\be
M = \frac{1}{n} \langle m \rangle= \frac{1}{n\beta} \frac{\partial \ln Z}{\partial B}=
\frac{\beta B} {\frac{3}{s(s+1)} - \beta c} = 
{B \, T_c \ov c (T-T_c)}\ .
\ee
So the system is in a linear paramagnetic phase with a magnetic susceptibility
\be
\mu = \frac{M}{B} = \frac{T_c}{c ( T - T_c )}\ .
\ee
The above perturbative calculation and results are valid as long as the magnetic field is not too strong.
The criterion is that the spin of states over which the integral in (\ref{Zferpert})
receives a substantial contribution, which is of order $j \sim \langle m \rangle = nM$,
not be larger than its range of validity $j \sim \sqrt{n \frac{s(s+1)}{3}}$ (see \eqn{devv}, \eqn{jsn}).
 This means that
\be
 B < \sqrt{\frac{s(s+1)}{n}} \, T\ .
\ee
So in effect the result for the linear magnetization is a small-$B$ result.

\subsection{Ferromagnetic regime}

For temperatures below the Curie critical temperature the perturbative evaluation fails, since large values
of $j$ are now dominating the partition function, and a fully nonperturbative calculation is needed.
We revert to the method of the previous section, using saddle-point thermodynamics in the large-$n$ limit.

\no
The partition function of the present system can be related to the partition function of uncoupled spins through a thermodynamic transformation. Returning to the large-$n$ formula (\ref{Zlargen}) and using the large-$n$ relation
$d_{s,n} (j) = - \partial_j D_{s,n} (j)$ (see (\ref{dD})) we have, after integrating by parts, that
\be
Z = \frac{\beta}{2\sinh \frac{\beta B}{2}} \int_{-\infty}^\infty dj ~\left(B + \frac{cj}{n}\right) ~
D_{s,n} (j)  ~~ e^{\beta B \, j + \beta \frac{c}{2n} j^2}\ .
\ee
The multiplicities $D_{s,n} (j)$ can be obtained from the nonperturbative result (\ref{Dnonpert})
in terms of the thermodynamic entropy of the uncoupled spins. The partition function is expressed as
\be
Z = \frac{\beta}{2\sinh \frac{\beta B}{2}}\int_{-\infty}^\infty dj  ~\sqrt{-\frac{\partial_j^2 S(j)}{2\pi}}
~ \left(B + \frac{cj}{n}\right)  ~~
 e^{S(j) + \beta B \, j + \beta \frac{c}{2n} j^2}\ .
\label{Znonp}
\ee
The factor in front of the exponential in the integrand is irrelevant in the large-$n$ limit.
Indeed, for typical values of $j$ of order $n$, all the terms in the exponent are of order $n$, while the
prefactor is of order $n^{-\half}$ and contributes to the exponent a term of order $\ln n$ which can be
omitted. The integral can be evaluated by the saddle point method
(again, omitting the subleading determinant factor). The saddle point condition is
\be
 S'(j) + \beta B + \frac{\beta c}{n} j = 0\ .
\label{sad}
\ee
The entropy $S(j)$ can be expressed in terms of the single spin free energy function $f_s ( \tau )$
through the Legendre transform (\ref{lege}).\footnote{To distinguish from the real temperature parameter $\beta$,
we renamed $\beta$ in (\ref{lege}) to $\tau$. In fact, $\tau$ plays the role of chemical potential for
$j$ and will be fixed by the saddle point method to minimize the free energy of the interacting system.}
We have
\be
f'_s (\tau ) = \frac{j}{n} ~,~~~~~
S(j) = n f_s (\tau ) - \tau j ~,~~~~
\partial_j S(j) = - \tau
\ee
and combining with the saddle point condition (\ref{sad}) we eventually obtain
\be
f'_s (\tau ) = \frac{\tau}{\beta c} - \frac{B}{c} = \frac{j}{n}\ .
\label{ferrosol}
\ee
The first equality above determines $\tau$, while the second one gives the dominant $j$ which
fixes the magnetization $\displaystyle \frac{j}{n}$ in the large-$n$ limit.

\no
Using the known form (\ref{ssp}) of $f_s (\tau )$ the equation for $\tau$ becomes
\be
\left( s+ {\textstyle \half} \right) \coth \left( s+ {\textstyle \half} \right) \tau - {\textstyle \half}
 \coth {\textstyle \frac{\tau}{2}} =
\frac{\tau}{\beta c} - \frac{B}{c}\ .
\label{transc}
\ee
This is a transcendental equation that does not admit an explicit solution. However, the properties
of the system are clear from its graphical representation. Fig. 1 depicts the situation for a
typical form of $f'_s (\tau )$ and three different values of $B$.
\begin{figure}
\centering
\includegraphics[width=0.4\textwidth]{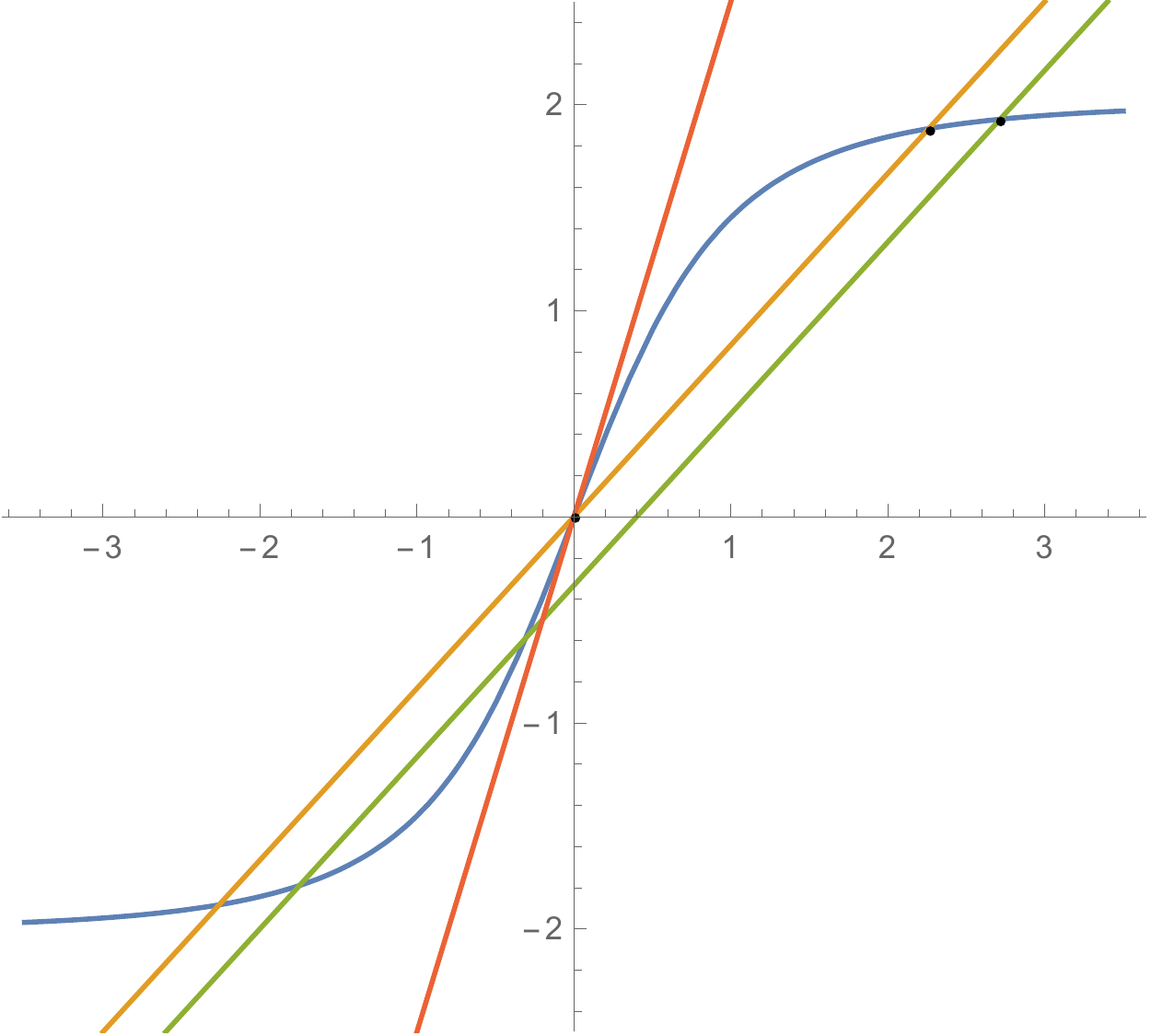}
\caption{Graphical solution of equation (\ref{transc}) for three cases: $T>T_c$, $B=0$ (red line), $T<T_c$, $B=0$ (orange line),
and $T<T_c$, $B>0$ (green line). Blue curve is $f'_s (\tau )$ and vertical axis represents magnetization.}
\end{figure}


\no
For $B=0$ there are either one solution for $\tau =0$ or three solutions at $0$ and $\pm \tau$,
depending on the slope $(\beta c)^{-1}$ of the straight line in the right hand side. From
\be
f''_s (\tau ) = -\left(\frac{s+ \half}{\sinh \left( s+ \half \right) \tau} \right)^2 +
\frac{1}{\left( 2\sinh \frac{\tau}{2} \right)^2}\ ,
\ee
the slope of the curve $f'_s (\tau)$ at $\tau=0$ is calculated as $f''_s (0) = \frac{s(s+1)}{3}$. Therefore, if
\be
\frac{1}{\beta c} > \frac{s(s+1)}{3}\ .
\ee
there will be only one solution at $\tau =0$, giving $j=0$. This is precisely the condition
for the temperature to be above the Curie temperature $T_c$ as given in (\ref{Cur}),
recovering the result that above the Curie temperature there is no spontaneous magnetization.

\no
For temperatures below the Curie temperature we have three solutions. The solution $\tau =0$,
however, is unstable as it corresponds to a local minimum of the partition function,
rather than a maximum. Indeed, the second derivative of the exponent in (\ref{Znonp}) is
\be
\partial_j^2 S(j) + \frac{\beta c}{n} = -\partial_j \tau + \frac{\beta c}{n}
=  - \frac{1}{\partial_\tau j} +  \frac{\beta c}{n} = -\frac{1}{n f''_s (\tau )} +  \frac{\beta c}{n} \ ,
\ee
where we used (\ref{ferrosol}) to put $j = n f'_s (\tau )$ in the last step.
From the known value of $f''_s (0)$ we see that at $\tau = 0$ the above will be negative
for temperatures above the Curie temperature, so $j=0$ is a maximum and we are in a paramagnetic
phase. For temperatures below $T_c$, however, the above becomes positive, indicating that
$j=0$ becomes a local minimum and is unstable. We are left with the two symmetric solutions
$\pm \tau$ giving a macroscopic magnetization $\frac{j}{n}$. For $B=0$
the system can in principle diffuse from one value of $m$ (that is, direction of the total spin) to any other
within this $j$, but the diffusion time is quadratic in $n$, and in the large-$n$ limit such transitions
are suppressed, signaling rotational symmetry breaking and a ferromagnetic phase.

\no
For nonzero $B$ the system's magnetization becomes aligned with $B$.
In the paramagnetic phase ($T > T_c$) the solution is unique, giving a magnetization increasing
from $0$ and asymptotically reaching the saturation value $\frac{j}{n} = s$ for strong magnetic fields. In the
ferromagnetic phase there are again three solutions. The middle one is unstable, but the
other two are stable under change of the magnitude of $j$ and inequivalent. The one with the lowest free energy should be
picked (the one maximizing (\ref{Znonp})), which is the one in the direction of $B$. (The opposite
one represents an equilibrium point of the macroscopic total spin pointing in the direction
opposite to $B$, unstable under global spin rotations.) For large enough $B$, this second
solution disappears.

\no
In conclusion, we demonstrated that the large-$n$ results of our analysis can be used to
derive the ferromagnetic phase transition and physical properties of the material.

\section{Generalization for several values of $s$}

The previous analysis can be generalized in several directions that are of interest
in various physical situations.
We list a couple of them with the corresponding results, starting from the case
of a mixture of spins.

If not all of the composed spins are the same, results change accordingly.
Clearly the order of composition of spins is immaterial, so only the total number
of each kind matters. In the partition function method, the
total partition function will be the product of individual ones
and the total number of spins
will be a sort of convolution of the individual numbers
$d_{s_i , n_i , j}$ for each kind of participating spin.

\subsection{Two kinds of spins}

In the simplest case of two kinds of spins, $n_1$ spins $s_1$ and $n_2$ spins $s_2$,
the partition function is
\be
Z = Z_{s_1}^{n_1} Z_{s_2}^{n_2}=  \left( \frac{x^{s_1 +1} - x^{-s_1}}{x-1} \right)^{n_1}
\left( \frac{x^{s_2 +1} - x^{-s_2}}{x-1} \right)^{n_2}\ .
\label{Z2}
\ee
A similar analysis as before, writing the denominators in (\ref{Z2}) as $(1-x^{-1})$ and expanding
in powers of $x^{-1}$, leads to the result
\begin{align}
&d_{s_1,n_1;s_2,n_2;j} = \frac{n_1 ! n_2 !}{(n_1 + n_2 -2)!} ~\times
\nonumber\\
& \sum_{a,b}
(-1)^{a+b} \, \frac{\Bigl( (s_1 +1)n_1 + (s_2 +1)n_2 - (2s_1 +1)a - (2s_2 +1)b -j-2 \Bigr)!}
{a! b! (n_1 -a)! (n_2 -b)! \left(s_1 n_1 + s_2 n_2 - (2s_1 +1)a - (2s_2 +1)b -j \right)!}\ ,
\label{opala}
\end{align}
where the summation is over all integer values of $a,b$
such that all factorial arguments are non-negative.
For $s_1 = s_2 =s$ \eqn{opala} becomes (\ref{opa}) for $n=n_1 + n_2$ upon using
a combinatorial identity.

\no
Similarly, the large-$n$ behavior (for both $n_1$ and $n_2$ large) is obtained by convoluting the
two $D_{s_1,n_1} (j)$ and $D_{s_2,n_2} (j)$ and then differentiating. The convolution
gives a Gaussian distribution with normalization $(2s_1 +1)^{n_1} (2s_2 +1)^{n_2}$ and variance
$\sigma^2 = n_1 \sigma_1^2 + n_2 \sigma_2^2$, namely
\be
D_{s_1,n_1;s_2,n_2,j}  \simeq D_{s_1,n_1;s_2,n_2}(j) = \frac{\sqrt{3}\, (2s_1 +1)^{n_1} (2s_2 +1)^{n_2}}
{\sqrt{2\pi s_1(s_1 +1)n_1  s_2(s_2 +1)n_2}}
\, e^{-\frac{3j^2}{2s_1(s_1 +1)n_1 + 2 s_2(s_2 +1)n_2}}\ .
\ee
Using \eqn{dD} with the above formula gives the desired $d_{s_1,n_1;s_2;n_2} (j)$.

\subsection{A distribution of spins}
Analogous formulae can be obtained for higher number of spin species, but they are rather tedious and not
terribly illuminating. A more interesting situation is a {\it large} distribution of various values of $s$, and its
behavior in the large-$n$ limit. Specifically, we consider a collection of spins, $n_s$ of them of value $s$,
for a total number $n = \sum_s n_s$. Such a situation could arise, for example, in the condensed matter
of mixed materials or in loop quantum gravity.

For a large number of spins we define their distribution
\be
\rho(s) = \frac{2 n_s}{n} ~,~~~ \sum_s \frac{n_s}{n} \simeq \int ds \rho(s) = 1\ ,
\ee
where the factor of $2$ accounts for the fact that the spacing between successive values of $s$ is $\half$.
Working in the partition function/generating function approach, the total ``free energy" of the system is
additive in the spins and is given by
\be
\ln Z = \sum_s n_s \ln Z_{s,1} (x) \simeq n \int ds \rho(s) \ln Z_{s,1} (x)\ .
\ee
For large $n$, the free energy will again be dominated by its saddle-point approximation and thus
the Central Limit Theorem will apply, reproducing a Gaussian distribution. Specifically we will have
\be
D_{n,j}  \simeq D_{n}(j) = \frac{N^n}{\sqrt{2\pi n} \, \sigma} \, e^{-\frac{j^2}{2n\sigma^2}}\ ,
\ee
with
\ba
&& N = \prod_s (2s+1)^{n_s /n} \simeq \exp\left\{\int ds \rho(s) \ln (2s+1)\right\}\ ,
\nonumber\\
&&
\sigma^2 = \sum_s \frac{n_s}{n} \, \frac{s(s+1)}{3} \simeq \int ds \rho(s) \frac{s(s+1)}{3}
\ea
and
\be
d_n (j) = -\frac{\partial D_n (j)}{\partial j} = \frac{N^n}{\sqrt{2\pi n^3} \, \sigma^3} \, j \,
e^{-\frac{j^2}{2n\sigma^2}}\ .
\ee

\no
There can be, on the other hand, situations in which the Central Limit Theorem does not hold, leading to
a more general distribution. For the Central Limit Theorem to hold we must have a situation where the second
moment (variance) of the total spin dominates in the distribution. Specifically, define
\be
\mu_{s,k} = \frac{1}{2s+1} \sum_{m=-s}^s m^k
\ee
and
\be
\mu_k = \sum_a \mu_{s_a , k} = \sum _s n_s \, \mu_{s,k}\ .
\ee
Then for the CLT to hold we must have
\be
\mu_{2k} \ll \mu_2^k
\ee
(odd moments vanish). Generically this is always true for a large distribution of spins. E.g., for $n$ equal spins,
both $\mu_2$ and $\mu_{2k}$ are of order $n$ so the left hand side is down by a power $n^{1-k}$.
The only possibility for the above to fail is if a macroscopic fraction of the total $\mu_2$ is contributed by
one, or few spins.

\no
To make this explicit, note that for large spins ($s$ $\gg$ 1), $\mu_{s,2k} \simeq (k+1)^{-1} s^{2k}$. For instance,
\be
\mu_{s,2} = \sigma_s^2 = \frac{s(s+1)}{3} \simeq \frac{1}{3} s^2
\ee
So for a large distribution of spins the above relation can be written as
\be
\frac{1}{(k+1)} \sum_a s_a^{2k} \ll \frac{1}{3^k} \left( \sum_a s_a^2 \right)^k\ .
\label{CLT}
\ee
Define
\be
w_a= \frac{s_a^2}{\sum_b s_b^2}\ .
\ee
The variables $w_a$ represent the fraction of the total variance that each spin contributes
and satisfy
\be
0 \le w_a \le 1 ~,~~~ \sum_a w_a = 1
\ee
In terms of $w_a$, relation (\ref{CLT}) can be written
\be
\sum_a w_a^k \ll \frac{k+1}{3^k}\ .
\label{cond}
\ee
Now assume that the maximal value of $w_a$ is $r$, that is $w_a \le r$. This implies
\be
w_a^k \le r^{k-1} w_a ~~~ \Longrightarrow ~~~ \sum_a w_a^k \le r^{k-1}\ .
\ee
Hence, if $r$ is very small, (\ref{cond}) will necessarily hold. The only possibility that it could
be violated is if $r$ is of order 1, which implies that the largest spin must
contribute a substantial fraction $r$ of the total variance. This could arise if, on top
of the smooth distribution of spins, there are a few large "outliers". Such a situation, e.g., would be a collection
of spins that follow a power law distribution
\be
s_a = \half \left[ \frac {2s n^\gamma}{a^\gamma} +\half \right] \sim s \left(\frac{n}{a}\right)^\gamma ~,
~~~\gamma >\half ~, ~~~ a = 1, \dots  , n\ .
\ee
This represents a distribution of $n$ spins, asymptotically reaching the value $s_a =s$, but with the largest spin being equal to $n^\gamma s$. (The "frills" of the shift by $\half$ and integer value are so that we obtain only integer and
half-integer spin values, with transitions between them at the point midway between two successive values.)
For such a distribution we see that the first spin contributes a fraction of the variance of order
\be
w_1 \sim \frac{1}{\zeta (2\gamma )}\ ,
\ee
with $\zeta ( \cdot )$ being Riemann's zeta function,
which is macroscopically big (e.g., for $\gamma = 1$, $w_1 \sim .6$). In such situations, the asymptotic
distribution has to be calculated on a case-by-case basis.

\section{(Anti-)symmetric composition of spins}

In the analysis so far we kept all irreducible components in the decomposition of spins, effectively considering individual spins as distinguishable.
We could, alternatively, consider spins that are identical and indistiguishable, either
bosonic or fermionic. In such cases we must keep only the totally symmetric part of the
decomposition, or the totally antisymmetric one. In more generality we could select
irreducible representations transforming under any specific mixed symmetry. This would correspond to
spins obeying ``parastatistics" or, more realistically, spins of ordinary statistics but possessing
a set of additional discrete internal degrees of freedom.

\no
It is clear that the counting of total spin components changes drastically in this case.
For example, the symmetric product of two spins with $s=\ha$ is a triplet, while the antisymmetric
product gives a singlet. As we shall see, the large-$n$ asymptotic behavior is also quite
different in this case.
We will consider here the case of (anti-)symmetric products of $n$ spins $s$.
The problem is trivial for $s=\half$. Then, there is a single symmetric component of spin $\frac{n}{2}$,
while there is a single antisymmetric spin-0 component for $n=2$ and no such components for
$n>2$. The situation becomes more interesting and nontrivial, however, for higher $s$.

\subsection{Partition function}

The qualitative change in the case of representations with specific symmetry is that we do not have any
obvious recursive relation in $n$ similar to (\ref{compo}), so the random walk approach is not available.
Hence, we resort to the partition function method. In fact, the {\it grand} partition function of the system,
where we weight particles with a chemical potential factor $e^{\mu n}$, is readily available:
it is the grand partition of a gas of non-interacting bosons (symmetric) or fermions (antisymmetric) in the energy
levels $-Bm$ with given temperature and chemical potential. Specifically
\be
{\cal Z}_s^\pm = \prod_{m=-s}^s \left( 1 \pm e^{\mu + \beta B m} \right)^{\pm 1}\ ,
\ee
where $-$ ($+$) corresponds to bosons (fermions). Writing $x=e^{\beta B}$ and
$y= e^\mu$, the above gives the generating function for the number of states with given
$z$-component, as usual
\be
{\cal Z}_s^\pm (x,y) = \sum_{n=0}^\infty \sum_{m=-s}^s D_{s,n,m}^\pm x^m y^n =
\prod_{m=-s}^s \left( 1 \pm y x^m \right)^{\pm 1}\ .
\label{pfpm}
\ee
We see that for bosons (symmetric) this is an infinite series in $y$, while for fermions (antisymmetric)
it is a
polynomial in $y$ of degree $2s+1$, signaling that we can have at most $2s+1$ spins composed
antisymmetrically, as expected. An expansion in $y$ gives the result for the symmetric ($-$) and
antisymmetric ($+$) fixed-$n$ generating function
\be
Z_{s,n}^+ (x) = \sum_{0 \leqslant k_1 < \cdots < k_n \leqslant 2s} x^{k_1 + \cdots + k_n - ns} \ ,\quad
Z_{s,n}^- (x) = \sum_{0 \leqslant m_1 \leqslant \cdots \leqslant m_n \leqslant 2s}^s x^{m_1 + \cdots + m_n -ns}\ ,
\label{sumk}
\ee
from which the generating function for the number of components with spin $j$ is obtained as usual
\be
\zeta_{s,n}^\pm (x) = \sum_j d_{s,n,j}^\pm x^j = (1-x^{-1} ) \, Z_{s,n}^\pm (x)\ .
\ee
We do not have explicit combinatorial expressions, analogous to (\ref{opa}), for generic spin $s$.
We have derived, nevertheless, a ``bosonization" formula that relates the results of the antisymmetric cases to those of the
symmetric ones, an exact expression for the above generating functions,
and a remarkable duality relation between $n$ and $s$. We present all these below.

\subsubsection{Bosonization formula}

In a standard method, we can relate $Z_{s,n}^+ (x)$ to $Z_{s,n}^- (x)$ by expressing the fermionic
excitation numbers $k_a$ in the sum (\ref{sumk}) in terms of bosonic ones $m_a$. Specifically, writing
\be
m_a = k_a - a +1\ ,
\ee
we see that the integers $m_a$ obey bosonic conditions $m_a \leqslant m_{a+1}$ and span the values
$0, \dots 2s-n+1$. In terms of the $m_a$ the fermionic ($+$) sum becomes identical to the bosonic ($-$)
one but with $s$ shifted by $-(n-1)/2$. Therefore
\be
Z_{s,n}^+ (x) = Z_{s+\frac{1 -n}{2},n}^- (x) ~~~~{\rm and~also}~~~~
D_{s,n,m}^+ = D_{s+\frac{1 -n}{2},n,m}^-\ .
\label{babos}
\ee
In particular, if $n$ exceeds $2s+1$ the partition function $Z^+$ vanishes, as expected. So the antisymmetric
result for $d_{s,n,m}^+$ is simply the symmetric one but for a reduced spin $s-\frac{n-1}{2}$. Hence we may
focus on either the symmetric or antisymmetric case, the other one being trivially related as above.

The above bosonization formula implies a corresponding relation between generalized generating functions that
also sum over spin. Specifically, define
\be
{\cal W}^\pm (x,y,w) = \sum_{s=0}^\infty w^{2s} \, {\cal Z}_s^\pm (x,y)\ .
\ee
Then the bosonization formula implies\footnote{
The additive constant $1$ below arises from the term $s$ $=$ $n$ $=$ $0$, 
which must be isolated before performing
shifts in the summation variables.}
\be
{\cal W}^- (x,y,w) = 1+w \, {\cal W}^+ (x,yw^{-1},w)\ .
\label{bosoW}
\ee
This relation is quite nontrivial to derive directly 
from the definition of ${\cal Z}^\pm$.

\subsubsection{Exact expression for the partition function}

The antisymmetric grand partition function can be written (writing, temporarily, $q$ instead of $x$
to evoke $q$-deformations) as
\be
{\cal Z}^+ (q,y) = \prod_{m=-s}^s \left( 1 \pm y q^m \right) = [1+y]_q^{2s+1}\ .
\ee
In the right hand side the $q$-deformed binomial appears, defined as
\be
[1+y]_q^N = \sum_{k=0}^N \frac{ [N]!_q}{[k]!_q [N-k]!_q} y^k\ ,
\ee
where the $q$-factorials are defined in terms of the $q$-deformation of (\ref{qdef})
\be
[n]!_q = \prod_{k=1}^n [k]_q = \prod_{k=1}^n \frac{ q^{\frac{k}{2}} - q^{-\frac{k}{2}}}
{q^{\half} - q^{-\half}}\ .
\ee
We obtain, then, for the $n$-spin antisymmetric generating function
\be
Z_{s,n}^+ (x) = \frac{[2s+1]!_x}{[n]!_x [2s+1-n]!_x} =
\prod_{k=1}^n \frac{[2s+2-k]_x}{[k]_x} =
\prod_{k=1}^{2s+1-n} \frac{[2s+2-k]_x}{[k]_x}\ .
\label{nf}
\ee
Using the bosonization formula we obtain the corresponding symmetric generating function
\be
Z_{s,n}^- (x) = \frac{[2s+n]!_x}{[n]!_x [2s]!_x} =
\prod_{k=1}^n \frac{[2s+k]_x}{[k]_x} =
\prod_{k=1}^{2s} \frac{[n+k]_x}{[k]_x}\ .
\label{nb}
\ee
Finally, using the explicit expressions for the $q$-deformed numbers we obtain
\be
Z_{s,n}^+ (x) = \prod_{k=1}^n \frac{x^{s+1} - x^{k-s-1}}{x^k -1} =
 \prod_{k=1}^{2s+1-n} \frac{x^{s+1} - x^{k-s-1}}{x^k -1}
\label{exf}
\ee
and
\be
Z_{s,n}^- (x) = \prod_{k=1}^n \frac{x^{s+k} - x^{-s}}{x^k -1} =
\prod_{k=1}^{2s} \frac{x^{\frac{n}{2}+k} - x^{-\frac{n}{2}}}{x^k -1}\ .
\label{exb}
\ee
Note that, although the above expressions for $Z_{s,n}^+$ and
$Z_{s,n}^-$ do not look manifestly related by bosonization, they can be shown to satisfy (\ref{babos})
by appropriate changes of the product variable $k$ in the numerator and denominator. We note the similarity
of these formulae to the partition function of some integrable spin chains \cite{Poly}.

The above formulae are remarkably compact, and an expansion in powers of $x$ would yield
combinatorial formulae for $D_{s,n}^\pm$ analogous to (\ref{opa}) for the case of no symmetry.
However, their
product form prevents us from obtaining such explicit expressions.
We also point out that the limit $x \to 0$ reproduces the total number of states for symmetric
($-$) or antisymmetric ($+$) products, as expected. In that limit, $[k]_x \to k$ and we obtain the standard
results
\be
N_{s,n}^- = Z_{s,n}^- (0) = \frac{(2s+n)!}{n! (2s)!} \ ,\qq
N_{s,n}^+ = Z_{s,n}^+ (0) = \frac{(2s+1)!}{n! (2s+1-n)!}\ .
\ee

\subsubsection{Duality relation}

We conclude by pointing out a duality relation between $s$ and $n$ in the
symmetric case, and an analogous one in the antisymmetric case. From the expressions in terms of
$q$-factorials in formulae (\ref{nf},\ref{nb}), we immediately see that
\be
Z_{s,n}^- (x) = Z_{\frac{n}{2},2s}^- (x) ~,~~~ D_{s,n,m}^- = D_{\frac{n}{2},2s,m}^-
\label{dualbos}
\ee
and that 
\be
Z_{s,n}^+ (x) = Z_{s,2s+1-n}^+ (x) ~,~~~ D_{s,n,m}^+ = D_{s,2s+1-n,m}^-
\label{hole}
\ee
This duality is, in fact, particle-hole symmetry in the antisymmetric (fermionic) case,
while bosonization maps it to a spin size-spin number duality in the symmetric (bosonic)
case. As a trivial example, the symmetric composition of $n$ spins-half ($s=\half$,
$n$) gives a unique irreducible component of spin $\frac{n}{2}$, as in ($s=\frac{n}{2}$, $1$).

These dualities can also be derived from the partition functions. For the antisymmetric one we have
\be
{\cal Z}_s^+(x,y) = \prod_{m=-s}^s (1+y x^m ) = \prod_{m=-s}^s [y x^m (1+ y^{-1} x^{-m} )]
= \prod_{m=-s}^s y x^m \prod_{m=-s}^s (1+ y^{-1} x^{-m} )\ .
\ee
Performing the first product, and changing $m$ to $-m$ in the second one, we obtain the relation
\be
Z_{s,n}^+ (x) = y^{2s+1} {\cal Z}_s^+ (x, y^{-1} )\ .
\ee
Expansion in powers of $y$ implies (\ref{hole}).
In terms of the spin generating function ${\cal W}^+$ the above implies
\be
{\cal W}^+ (x,y,w) = y \, {\cal W}^+ (x,y^{-1},yw)\ .
\ee
Combining the above formula with the bosonization formula (\ref{bosoW}) yields simply
\be
{\cal W}^- (x,y,w) = {\cal W}^- (x,w,y)\ ,
\ee
which, upon expanding in powers of $y$ and $w$ gives the duality relation (\ref{dualbos}).

We also notice an "inversion" relation between the symmetric and antisymmetric cases, evident from the
first product in the expressions (\ref{exf}) and (\ref{exb}): they are related by the formal mapping $s \to -s-1$:
\be
Z_{s,n}^- (x) = (-1)^n \, Z_{-s-1,n}^+ (x)\ ,
\ee
which is, in fact, completely analogous to the antisymmetry we introduced in the random walk approach.
In a sense, the antisymmetic spin product corresponds to a reversal of the sign of the numbers of
spin states $2s+1 \to -2s-1$. The full implication of this observation, and its possible connection to
a random walk approach, are not yet known.

\subsection{Asymptotics for large $s$ and $n$}

The above duality formulae demonstrate that a large-$n$ analysis makes sense only when
we also take the large-$s$ limit. Indeed, keeping $s$
small in the symmetric case while increasing $n$ is, by duality, the same as considering $2s$ spins in a
large, spin-$n/2$ representation. In this section
we perform the full large-$s$, large-$n$ analysis. As we will see, the scaling properties of this
system are quite different from the ones for distinguishable spins.

\subsubsection{Leading behavior}

We will consider the symmetric case, the antisymmetric one being related through bosonization.
The starting point will be the generating function (\ref{exb}). In the present case it is no longer the
$n$-th power of a function, but the saddle point method is still applicable.

The number
of states with $z$-component equal to $m$ is obtained in the standard way
\be
D_{s,n,m} =  \frac{1}{2\pi} \int_0^{2\pi} Z_{s,n}^- \, (e^{it} ) e^{-imt} dt\ ,
\ee
where, from the second form of $Z_{s,n}$ in (\ref{exb}), we have
\be
Z_{s,n}^- (e^{it} ) = \prod_{k=1}^{2s} \frac{\sin\frac{n+k}{2} t}{\sin\frac{k}{2} t}\ .
\ee
For large values of $s$ and $n$ this integral will be dominated by small values of $t$ (this
will be justified in the sequel). The log of the generating function ("free energy") is
\be
\ln Z_{s,n}^- (e^{it} ) = \sum_{k=1}^{2s} \left( \ln \sin\frac{n+k}{2} t - \ln \sin\frac{k}{2} t \right)\ .
\ee
Expanding the summand in powers of $t$ we have
\be
\ln Z_{s,n}^- (e^{it} ) = \sum_{k=1}^{2s} \left[ \ln\frac{n+k}{k}
+ \sum_{\ell=1}^\infty c_{2\ell} \left( (n+k)^{2\ell} - k^{2\ell} \right) t^{2\ell} \right]\ .
\label{sumk}
\ee
The $c_{2\ell}$ are fixed numerical coefficients, all of them negative:
\be
c_{2\ell} = \frac{(-1)^\ell \, B_{2\ell}}{2\ell\, (2\ell)!}\ ,
\ee
with $B_{2\ell}$ being the Bernoulli numbers. The coefficient of the quadratic term, in particular,
is $c_2 = -\frac{1}{24}$.

Exchanging the sums, the summation over $k$ can be performed in each term in the expansion.
For $2s$ large, the sum can be replaced by an integral, the error being of subleading order. We obtain
\be
\ln Z_{s,n}^- (e^{it} ) = \ln\frac{(n+2s)!}{n! (2s)!}
+ \sum_{\ell =1}^\infty
\frac{c_{2\ell}}{2\ell+1} \left( (n+2s)^{2\ell+1} - n^{2\ell+1} - (2s)^{2\ell+1} \right) t^{2\ell}\ .
\label{terms}
\ee
The main point of the saddle-point argument is the observation that, when {\it both} $n$ and $s$
become large, then the leading quadratic term
dominates in the expansion. To demonstrate this, we will consider the three possible scaling
cases: (i) $n$ and $s$ scale in the same way; (ii) they both grow, but one of them grows faster than the other;
and (iii) one of them remains finite while the other one grows.

We start with case (i), that is, $n,s \to \infty$ but with $s/n$ remaining finite. In this case, the
coefficient of the term $t^{2\ell}$ is of order $n^{2\ell +1}$, since all terms in it are of that order.
This implies that the function is
dominated by the quadratic therm. To make this explicit, define the rescaled variable ${\tilde t} = n^\frac{3}{2} t$.
In terms of $\tilde t$ the expression (\ref{terms}) becomes
\be
\ln Z_{s,n}^- = \ln\frac{(n+2s)!}{n! (2s)!}
+ \sum_{\ell =1}^\infty n^{1-\ell}\,
\frac{c_{2\ell} }{2\ell+1} \left( (1+2s/n)^{2\ell+1} - 1 - (2s/n)^{2\ell+1} \right) {\tilde t}^{\, 2\ell}
~,~~~{\tilde t} = n^\frac{3}{2} \, t
\ee
In the new variable, the terms scale as $n^{1-\ell}$.
So all the terms with $\ell >1$, that is, higher than quadratic, become negligible in the scaling limit and
can be dropped. This also tells us that the generating function will have non-negligible values when ${\tilde t}$
is not much bigger than 1, that is, when $t$ is of order $n^{-\frac{3}{2}}$, which justifies our claim that the
generating function is dominated by small values of $t$. Keeping only the quadratic term, and using the
value $c_2 = -\frac{1}{24}$, the free energy becomes
\be
\ln Z_{s,n}^- (e^{i t} ) = \ln \frac{(n+2s)!}{n! (2s)!} -\frac{n s (2s+n)}{12} t^2\ ,
\ee
giving in that limit a Gaussian generating function of the form
\be
Z_{s,n}^- (e^{i t} ) = \frac{(n+2s)!}{n! (2s)!} e^{-\half \sigma_-^2 t^2} ~,\quad 
\sigma_-^2 = \frac{n s (2s+n)}{6} \ .
\ee
Its Fourier transform will give a normal distribution for  $m$ with variance $\sigma_-^2$:
\be
D_{s,n}^- (m) = \sqrt{\frac{3}{\pi n s (2s+n)}} ~\frac{(n+2s)!}{n! (2s)!} ~e^{-\frac{3}{n s (2s+n)} m^2}\ .
\label{Dlargens}
\ee

The situation is similar in case (ii) where $n$ and $s$ have different scalings. Assuming $n$ scales faster than $s$
(the opposite situation being equivalent, due to duality), this means  $n,s \to \infty$ but $s/n \to 0$. In this case,
keeping the leading contribution in the expression of the coefficients, we see that the coefficient of the term $t^{2\ell}$
is of order $n^{2\ell} s$. Again, defining the new variable ${\tilde t} = n s^\half t$, the function becomes
\be
\ln Z_{s,n}^- = \ln\frac{(n+2s)!}{n! (2s)!}
+ 2 \sum_{\ell =1}^\infty s^{1-\ell}~  c_{2\ell} \, {\tilde t}^{\, 2\ell}
~,~~~{\tilde t} = n s^\half \, t\ .
\ee
Since $s$ is large (even though subleading to $n$), all terms higher than quadratic are negligible. We again
recover a normal distribution for $m$, but now with a variance (given by the coefficient of $t^2$)
\be
\sigma_-^2 = \frac{s n^2}{6}\ ,
\ee
which is, in fact, the same as the one in case (i) when we keep the leading contribution ($2s+n \sim n$).

Finally, we consider case (iii) where only $n$ increases but $s$ remains finite (or vice versa). In this case
we cannot approximate the sum over $k$ in (\ref{sumk}) with an integral. However, given that the
term $n^{2\ell}$ in the expansion of $(n+k)^{2\ell}$ in the only leading one in the coefficients,
it can be easily summed to give $2s n^{2\ell}$ and the scaling of the coefficient is as in case (ii).
In this case, however, the higher coefficients are {\it not} subdominant, since, even after redefining
${\tilde t} = n s^\half t$ to make the quadratic coefficient of order 1, the higher coefficients do not
scale down, as $s^{1-\ell}$ is no longer small.

We can also understand the fact that both $s$ and $n$ have to be taken large in order to have a scaling
limit as a criterion for validity for the normal distribution of $m$. The maximal value of $|m|$ is
obviously $n s$, happening when all spins have the value $+s$ or $-s$. For the normal distribution to
be a good approximation, the spread of $m$ must be much smaller than the cutoff value $n s$. So
we must have
\be
\sigma_- \ll n s ~~\Rightarrow ~~ \frac{n s (2s+n)}{6} \ll n^2 s^2 ~~~~{\rm or,} ~~~~
\frac{1}{n} + \frac{1}{2s} \ll 1\ ,
\ee
which means that both $n$ and $s$ must be large, irrespective of their relative size, with the
smaller of them determining the quality of the scaling approximation.

So we see that the large-$n$ large-$s$ scaling of the symmetric case is in every case a normal distribution for the $z$ component
of the total spin, as given in (\ref{Dlargens}). This distribution is invariant under the duality transformation
exchanging $n$ and $2s$, as expected.

The corresponding distribution for the antisymmetric case can be obtained
by bosonization, shifting $s$ to $s-\frac{n-1}{2}$
\be
D_{s,n}^+ (m) = \sqrt{\frac{3}{\pi n s (2s-n)}} ~\frac{(2s+1)!}{n! (2s+1-n)!} ~e^{-\frac{3}{n s (2s-n)} m^2}\ .
\ee
In the above we kept the exact bosonization shift in the (large) prefactor, to ensure that the total number of states
is correctly reproduced, but dropped subleading terms in the expression for $\sigma^2$.

The same result is obtained by applying the formal mapping $s \to -s-1$, as pointed out in the last section,
to the variance of the symmetric result. Clearly the criterion for proper scaling in the antisymmetric case becomes
$2s-n \gg 1$, $n \gg 1$.

The calculation of the number of spin $j$ components proceeds as usual. $d_{s,n} (j)$ is given by
the derivative of the distribution for $m$ as in (\ref{dD}). The spin $j=J_{sm}$ that maximizes it is the
same as the standard deviation of $m$, that is
\be
J_{s,n}^- = \sqrt{\frac{n s (2s+n)}{6}} ~,~~~ J_{s,n}^+ = \sqrt{\frac{n s (2s-n)}{6}}\ .
\ee

In summary, we see that the distribution of spins is broader than the distinguishable case when we
consider symmetric spins, and narrower when we consider antisymmetric ones. Indeed, for $s,n$ large,
the three variances are
\be
\sigma^2_\pm = \frac{ns^2}{3} \mp \frac{n^2 s}{6}  ~,~~~ \sigma^2 =  \frac{ns^2}{3}
\ee
and therefore are related as
\be
\sigma^2_+  < \sigma^2 < \sigma^2_- ~,~~~ \sigma_-^2 + \sigma_-^2 = 2 \sigma^2\ .
\ee
They become equal when $s \gg n$, as expected. Indeed, when the composed spins are
classical ($s \gg 1$) and relatively few in number ($n \ll 2s+1$) their distribution remains essentially
classical and their statistics are irrelevant.

\subsubsection{Subleading and nonperturbative corrections}

We conclude by pointing out that we can also calculate subleading corrections to the large-$n,s$
distribution. The calculation proceeds similarly as the one for distiguishable spins. In the present
case, in addition to including the quartic term in $t$ in the expression for $\ln Z_{s,n} (e^{it} )$, we must also
perform the exact summation over $k$ in the quadratic term, rather that the integral, since their difference
contributes a subleading term. In fact, this modifies the coefficient of the quadratic term to
\be
\sigma_-^2 = \frac{n s (2s+1+n)}{6} ~,~~~~~ \sigma_+^2 = \frac{n (s+1) (2s+1-n)}{6}\ .
\ee
These expressions have the advantage of being dual to each  other under both the exact bosonization transformation
$2s \to 2s +1 -n$ and the inversion transformation $2s+1 \to -2s -1$,
rather than their large-$n,s$ versions. Using the expression
in (\ref{terms}) and the value of $c_4 = -\frac{1}{2880}$ to include the quartic term, we obtain overall
for the symmetric case
\be
D_{s,n}^- (m) =\sqrt{\frac{3}{\pi n s (2s+n)}} ~\frac{(n+2s)!}{n! (2s)!} ~ e^{-\frac{1}{12}\delta_- } ~
e^{-\frac{3}{n s (2s+1+n)} (1-\delta_- ) \, m^2 - \epsilon_- \, m^4}\ ,
\ee
where
\be
\delta_- = \frac{3(n^2 + 2ns + 4s^2)}{5ns(2s+n)} ~,~~~~
\epsilon_- = \frac{9(n^2 + 2ns + 4s^2)}{10 n^3 s^3 (2s+n)^3}\ .
\ee
Since $m^2 \sim \sigma_-^2 \sim ns(2s+n)$, we see that both corrections are of order $\delta_-$,
which is of the same order as the parameter controlling the scaling limit, that is,
\be
\delta_- \sim \frac{1}{n} + \frac{1}{s} \sim \max \left( \frac{1}{n}\, ,\, \frac{1}{s} \right)\ .
\ee
The expression of $\delta_+$ and $\epsilon_+$  for the antisymmetic case can be obtained by bosonization,
$s \to s-\frac{n}{2}$, or, more simply, $s \to -s$.

Finally, nonperturbative corrections are in principle obtainable through the Legendre inversion
of $\ln Z_{s,n}^\pm$ to $S_{s,n}^\pm$. Due to the complicated form of the partition function, however,
this cannot be done explicitly. Assuming $s,n \gg 1$, expressing sums as integrals and changing variables,
the equation relating $m$ and $\beta$ in the symmetric case can be brought to the form
\be
\frac{1}{\gamma (1-\gamma )} \left( \int_0^\gamma - \int_{1-\gamma}^1 \right)
d\tau \, \tau  \coth [(2s+n)\beta \tau] = - \frac{m}{ns} ~,~~~~
\gamma := \frac{2s}{2s+n}\ .
\ee
The integral can be expressed in terms of the dilogarithm function, but the inversion to $\beta (m)$
cannot be done explicitly. As a check of the validity of the procedure, note that for
$\beta = \pm \infty$, $  \coth [(2s+n)\beta t] = \pm 1$
and the integral becomes elementary, giving $m = \mp ns$, the expected extreme values.

From the scaling of the above equation we see that for $\frac{m}{ns}$ of order 1, $1/\beta$ is of
order $2s+n$; that is, due to the modified scaling properties of the (anti)symmetric system, temperatures
of order $2s+n$ are required to excite a macroscopically big number of states (as opposed to temperatures
of order 1 in the distinguishable case). The entropy, nevertheless, turns out to be of order $2s+n$, as expected.

\section{Conclusions}

In conclusion, we have analyzed the multiplicities of the decomposition of an arbitrary number of $SU(2)$
representations (spins) into irreducible components in various situations involving identical spins, a mixture of spins or
a statistical distribution of spins, as well as indistinguishable spins with total symmetry or antisymmetry, and
derived results for the asymptotic behavior of these multiplicities in the limit of a large number of spins.
Our derivations did not rely on group theory results such as characters, but used quite
elementary techniques, invoking concepts of partition functions and random walks.
A specific physical example of coupled spins exhibiting ferromagnetism was worked out in detail using our methods.

There are various other potential applications or directions for generalization of the results.
In terms of applications, the results for the multiplicities have been used in studies of the completeness of the Bethe ansatz
for higher spin chains \cite{Nepo}. Further, as was pointed out in \cite{CuZa} and the references cited there,
the results for the number of singlet components ($j=0$) have been relevant in elasticity theory, quantum chemistry 
and nuclear physics, since they alre represent the
number of linearly independent three-dimensional isotropic rank-$n$ tensors. More generally, any system involving many
$SU(2)$ components is a candidate application. Other than spin materials, such situations arise in some matrix model
and brane configurations as well as in loop (random lattice) quantum gravity and black holes.

An interesting physical application of our results for symmetric (bosonic) composition would be a
Bose condensate of
molecules with spin \cite{StUe}. In that case, since the spatial wavefunction is identical for all the molecules,
the spins are in a
fully symmetric state realizing the situation analyzed in section 6.  In general, not much is
known about the magnetic
properties of such Bose condensates. Their behavior, in the presence also of mutual $SU(2)$-invariant
interactions,
could be examined using our large-$n$ results for the symmetric case.

Finally, we point out that there are several possible generalizations of our analysis: we could look at mixed,
nonabelian symmetries of spin composition, treating the spins as parabosons, parafermions or "quarks", 
compositions of irreps of higher rank groups, such as $SU(N)$, etc.
Moreover, the
situation of deformed $SU(2)_q$ spins could be examined, with the corresponding $q$-comultiplication rule
for spin composition, in the case of $q$ being a root of unity when the multiplicities of representations differ
form the standard $SU(2)$ case. These applications and generalizations are the suject of further investigation.

\section*{Acknowledgments}

The authors would like to thank Itzhak Bars for useful comments, Cosmas Zachos
for correspondence and discussion of his work with T. Curtright, and Pouyan Ghaemi for conversations
and insights on the condensed matter properties of Bose condensates.
The research of A.P. is supported by NSF grants No.~1213380 and 1519449 and by a PSC-CUNY grant.

\end{document}